# A FINITE STRAIN MODEL OF STRESS, DIFFUSION, PLASTIC FLOW AND ELECTROCHEMICAL REACTIONS IN A LITHIUM-ION HALF-CELL

A.F. Bower[1], P.R. Guduru and V.A. Sethuraman

School of Engineering, Brown University, Providence RI 02912, USA.

## Abstract

We formulate the continuum field equations and constitutive equations that govern deformation, stress, and electric current flow in a Li-ion half-cell.  The model considers mass transport through the system, deformation and stress in the anode and cathode, electrostatic fields, as well as the electrochemical reactions at the electrode/electrolyte interfaces.  It extends existing analyses by accounting for the effects of finite strains and plastic flow in the electrodes, and by exploring in detail the role of stress in the electrochemical reactions at the electrode-electrolyte interfaces.  In particular, we find that that stress directly influences the rest potential at the interface, so that a term involving stress must be added to the Nernst equation if the stress in the solid is significant. The model is used to predict the variation of stress and electric potential in a model 1-D half-cell, consisting of a thin film of Si on a rigid substrate, a fluid electrolyte layer, and a solid Li cathode.  The predicted cycles of stress and potential are shown to be in good agreement with experimental observations.

## 1. Introduction

Mechanical degradation under cyclic charging and discharging is one of several factors that limit the life of modern Li-ion batteries (Aurbach *et al.* 2002, Vetter *et al.* 2005, and Zhang *et al.* 2001). Mechanical degradation can occur in materials used for both the anode and cathode, but the problem is particularly severe in batteries that use either pure Si, or a composite in which Si is one of the constituents, for the anode.  Si is an attractive choice for use as a negative electrode because its charge capacity exceeds that of graphite by nearly an order of magnitude [Tarascon and Armand (2001), Li and Dahn (2007), Obrovac and Christensen (2004), and Sethuraman *et al.* (2010a)], and so has the potential to increase the capacity and specific energy density of a cell by as much as 35%.  However, the anode must accommodate a volumetric strain of as much as 270% in the Si when fully charged.  This strain generates substantial stresses, which generally lead to mechanical failure after only a few cycles of charge.  A wide range of candidate Si based anodes have been explored in recent years in an attempt to develop a robust microstructure [Kasavajjula *et al.* (2007) and Chan *et al.* (2008)].  However, while many designs show promise, a commercial lithium-ion battery with composite Si anode is not yet available.

---

The need to understand the failure processes in batteries has motivated a number of recent models of stress evolution, diffusion, and fracture in representative battery microstructures. These include detailed models of stress evolution in a single particle [Verbrugge and Cheng (2008, 2009) and Cheng and Verbrugge (2009, 2010)], nanowires [Deshpande *et al.* (2010)], or thin films [Bhandakkar *et al.* (2010) and Haftbaradaran *et al.* (2010)]; models of stress induced phase transitions in cathode materials [Meethong *et al.* (2007), Tang *et al.* (2009)] as well as more detailed models which consider collective behavior of an entire microstructure [Garcia *et al.* (2005), Christensen and Newman (2006), Wang and Sastry (2007), Golmon *et al.* (2009), Renganathan *et al.* (2010) and Christensen (2010)].

Existing studies of stress evolution in battery materials have two limitations. Firstly, nearly all existing studies treated the mechanical deformation in the solid using the linear theory of elasticity. This is a reasonable approximation for graphitic anodes, where the volumetric strains are relatively small, but is likely to lead to significant errors for materials such as Si, which experience large volumetric strains. A recent study by Christensen and Newman (2006) is a notable exception, but their treatment of finite deformations differs from the approach normally used in finite elasticity and plasticity. In addition, recent controlled experimental measurements of stress in Si films during cyclic lithiation and delithiation [Sethuraman *et al.* (2010c)] have demonstrated that the Si does not remain elastic, but deforms plastically during every cycle. Plastic flow is likely to significantly accelerate damage, and the experimental data reported by Sethuraman *et al.* (2010b) suggests that the energy dissipated in plastic flow can be a significant fraction (as much as 40%) of the total energy loss in the cell, so it is of interest to extend existing models to account for its effects.

Secondly, the measurements reported by Sethuraman *et al* (2010c) have shown that stress in the electrode has a direct effect on the open-circuit electric potential of the cell. A compressive stress of 1GPa in the film was observed to reduce the open-circuit potential difference across the half-cell by 100-125 mV. This coupling is to be expected on thermodynamic grounds: for example, plastic flow in the film dissipates energy, and so must reduce the electrical energy that can be extracted from the cell. This can only occur if the stress and electric potential are coupled in some way. Sethuraman *et al.* (2010c) gave a more rigorous thermodynamic argument along these lines using a small-strain model. They predicted *ca.* 65 mV reduction in open-circuit potential for a 1GPa stress, in reasonable agreement with measurements. Existing models of stress evolution in battery microstructures only include an indirect coupling between stress and electric potential. The reactions are typically modeled using a Butler-Volmer equation, in which the reaction rate is a function of the potential difference across the interface, and may also depend on the concentrations of the reacting species adjacent to the interface. In these models, stress can only influence the interface reaction by altering the concentrations of the reacting species. The measurements and calculations reported by Sethuraman *et al* (2010c) indicate that applying a stress to an electrode at *fixed* concentration will have a measurable effect on the open-circuit potential of the cell (or equivalently, on the rate of reaction at a fixed overpotential).

Our goal in this paper is to address these limitations. We have three specific objectives: first, to present a rigorous derivation of the field equations governing coupled stress, diffusion, electric fields, and electrochemical reactions in an electrochemical cell; second, to extend existing models deformation in battery microstructures to account for the effects of finite strains and plastic flow in the deformable solids; and third, to establish the role played by stress in driving the electrochemical reactions. We emphasize that our focus is on the role of deformation and mechanical stress in high-capacity battery anode materials, and on the coupling between stress and electrochemistry. We do not aim to construct a complete and detailed model of a realistic electrochemical cell. For example, although our model includes a description of the electrolyte, we have purposely adopted the simplest possible description that will ensure charge is conserved properly in our model system. Furthermore, a range of phenomena occur in battery materials that are not considered here. These include formation of solid-electrolyte-interphase layer on the anode [Peled (1979)], phase transformations, particularly in cathode materials [Meethong *et al.* (2007, 2008)], which have been modeled extensively using phase field methods [Burch and Bazant (2009), Tang *et al.* (2009), and Singh *et al.* (2008)].

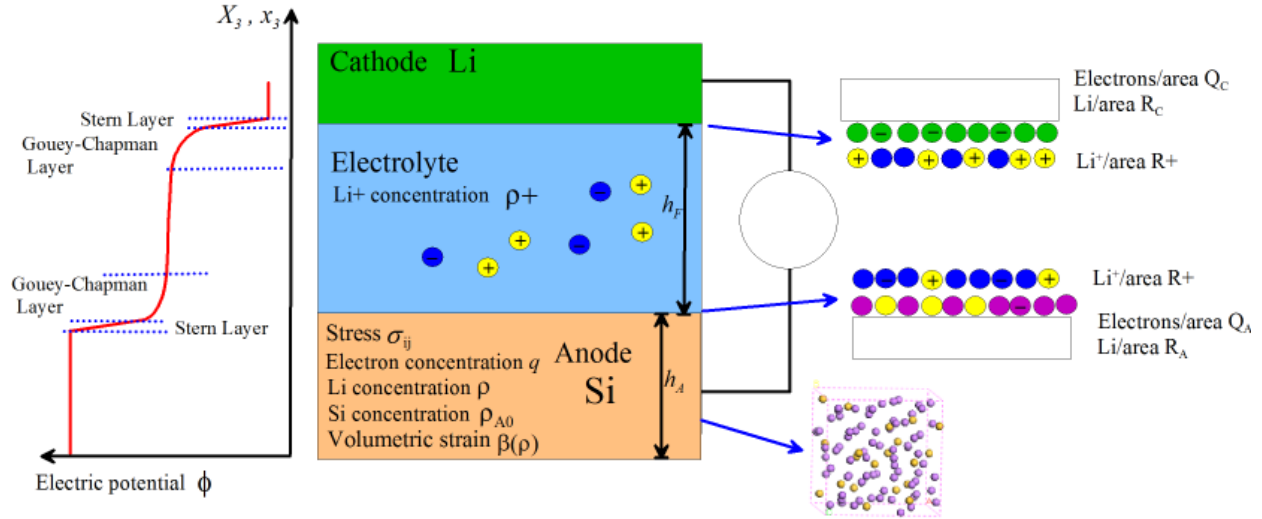

*Figure 1: An idealized model of a Li-ion half-cell, consisting of a Si anode, a solid Li cathode, and a fluid electrolyte.*

It is helpful to begin with a brief summary of the operation of a half-cell. Fig. 1 shows a schematic of a representative Li-ion half-cell, which consists of a solid anode, a fluid electrolyte, and a solid cathode.[2] The anode and cathode are represented as planar films for simplicity, but in reality the geometry is arbitrary. The anode and cathode are both deformable solids, while the electrolyte is a fluid. The entire assembly is subjected to both electrical and mechanical loading. These cause the anode, cathode and electrolyte to deform; induce a flow of electrons and Li (or Li-ions) through the system; and cause electrochemical reactions near the electrode/electrolyte interfaces. Our goal is to formulate the equations governing these processes.

When the cell is first assembled, and left at open circuit, electrochemical reactions occur at the electrode/electrolyte interfaces, which continue until the system reaches equilibrium. At the cathode, Li dissolves in the electrolyte through the reaction $Li \rightarrow Li^+ + e^-$. As a result, the cathode accumulates a negative charge, while a positive charge develops in the electrolyte. The charges are localized around the interface between electrode and electrolyte, causing an intense electric field to develop in a region (known as the 'Stern Layer') with atomic-scale dimensions adjacent to the interface, together with a more diffuse region of charge (known as the Gouy-Chapman layer) in the electrolyte (Hamann *et al.* 1998). A similar process occurs at the anode. Here, Li-ions in the electrolyte combine with electrons in the anode through the reaction $Li^+ + e^- \rightarrow Li$ and Li atoms alloy with the anode. The anode develops a positive charge, while the electrolyte develops a negative charge, establishing a double-layer potential at the electrode-electrolyte interface. The Li forms an interstitial solution in the anode, causing it to increase in volume, and inducing stresses in the anode. The reactions continue until the electric potential difference across the electrode/electrolyte interfaces prevents further charge transfer across the interface. At open-circuit equilibrium, the anode has a small positive charge, and an initial concentration of Li in solid solution. The electrolyte has a concentration of $Li^+$ in solution, while the cathode has a small negative charge. At this point the electrical potential of the anode $\phi_A$ exceeds that of the cathode, $\phi_C$ by the open circuit potential.

---

[2] Si is typically referred to as 'anode' in a full-cell configuration, and the same terminology is used in the half-cell configuration studied here. The electric potential of the Si electrode exceeds that of the Li reference electrode in a half-cell, however.

If an electrical connection is made between the anode and cathode, current flows between them, permitting the electrochemical reactions to continue. Several processes accompany the reactions. Electric currents flow through both the cathode and anode. Li atoms intercalate into the anode and diffuse through it, driven by variations in chemical potential. Phase changes may occur in the anode material: for example, a crystalline Si anode transforms to an amorphous solid during the first cycle of charge (Limthongkul *et al.* 2003). The anode also swells as the internal concentration of Li increases, and if the solid is constrained, substantial stresses will develop. The stresses are sufficient to cause plastic flow as shown by Sethuraman *et al.* (2010b, 2010c). A pressure may also develop in the electrolyte, which may induce stresses in the solid Li cathode. Our goal is to derive the equations governing all these processes.

The focus of this paper is the various *mechanical* processes that accompany the electrochemical reactions. In particular, we shall derive field equations and constitutive relations governing deformation, diffusion and electric current flow through the anode, and establish the role played by mechanical stresses in the electrochemical reactions. The mechanical processes cannot be treated in isolation, however. A consistent and rigorous model must also account for the electric fields in the system; and must also include at least a rudimentary model of the electrolyte.

To this end, we treat the entire cell as an isolated thermodynamic system, consisting of a solid anode, a fluid electrolyte, and a solid cathode. The electrochemical reactions are assumed to take place in two thin layers at the electrode/electrolyte interfaces. Each part of the system, including the reaction layers, is modeled explicitly.

The remainder of this paper is organized as follows. In the next section, we summarize the kinematic relations and conservation laws that govern deformation, diffusion, and electric fields in each part of the cell, treating the anode, cathode, and reaction layers in turn. We then derive an expression for the rate of change of free energy in the system, which is used to determine equilibrium conditions and thermodynamically consistent constitutive equations governing deformation, electric fields, reactions, and diffusion in the system. The results are applied to predict the evolution of stress, concentration and electric potential in the model cell tested experimentally by Sethuraman *et al* (2010c), and the predictions of the model are compared to their experimental data.

## 2. Kinematics and conservation laws

We begin by formulating the various conservation laws and deformation measures that describe the various parts of the cell.

*2.1 The Anode*

It is convenient to introduce the unstressed configuration of the anode with zero concentration of Li as a 'reference configuration' for describing shape changes and transport. The reference configuration occupies a volume $V_{0A}$, and has external normal $N_i$. Part of the boundary of the solid, (denoted by $A_{0Aext}$ on the undeformed surface, or $A_{Aext}$ on the deformed surface) is subjected to a traction (force per unit deformed area) $t_i$, while a second portion ($A_{0Ae}$ on the reference configuration, or $A_{Ae}$ in the current configuration) is in contact with the electrolyte.

Following Larché and Cahn (1978), we idealize the anode as a network solid, with molar density $\rho_{0A}$ per unit reference volume of lattice sites. Li will be assumed to form an interstitial solution in this network. The molar density of Li per unit reference volume in the anode will be denoted by $\rho$. We may then define the molar concentration of Li in the anode as $c = \rho / \rho_{0A}$. The concentration of Li$^+$ in the anode will be neglected. In addition, the network lattice material (Si or C) is taken to be insoluble in the electrolyte.

The Li may diffuse through the interior of the anode, driven by variations in concentration or stress. The resulting molar flux of Li crossing unit reference area will be denoted by $j_i$. Li mass must be conserved during this process, which requires

$$\frac{d}{dt}\int_{V_{0A}} \rho \, dV + \int_{S_{0A}} j_i N_i \, dA = 0 \tag{1}$$

for any sub-volume $V_{0A}$ bounded by a surface $S_{0A}$ within the undeformed anode. Applying the divergence theorem, and localizing leads to the conservation law

$$\dot{\rho} + \frac{\partial j_j}{\partial X_j} = 0 \tag{2}$$

where the derivative is taken with respect to position in the undeformed solid. Li may also flow out of the anode and into the layer where the electrochemical reaction occurs. The molar flux of Li per unit reference area across $A_{A0e}$ will be denoted by $j_{Ae}$. Li conservation at the exterior surfaces of the anode requires

$$j_i N_i = 0 \quad X_i \in A_{0Aext} \qquad j_i N_i = j_{Ae} \quad X_i \in A_{0Ae} \tag{3}$$

In addition to Li, the anode contains a distribution of free electrons, which may flow through the anode, and participate in the electrochemical reaction at the interface. We denote the molar density of these free electrons per unit reference volume in the anode by $q$, and the flux of electrons crossing unit area of internal surface within the reference solid by $i_i$. Conservation of electrons requires that

$$\dot{q} + \frac{\partial i_j}{\partial X_j} = 0 \tag{4}$$

Electrons may also flow across the external surfaces of the anode, pass through an external circuit, and reach the cathode. This exchange results in an flux density $i_{Aext}$ per unit undeformed area flowing across $A_{0Aext}$. The current is positive if electrons flow out of the anode. Electrons may also flow out of the solid anode across the anode/electrolyte interface $A_{0Ae}$, and into the interfacial region at the electrolyte/anode interface where the electrochemical reaction occurs. The flow of electrons out of the solid anode per unit reference area will be denoted by $i_{Ae}$. Conservation of electrons at external surfaces and interfaces of the anode then gives the boundary conditions

$$i_i N_i = i_{Aext} \quad X_i \in A_{0Aext} \qquad i_i N_i = i_{Ae} \quad X_i \in A_{0Ae} \tag{5}$$

Finally, the anode may also change its shape, either as a result of external mechanical loading, or as a result of Li insertion. During this process, a generic lattice point with position $X_i$ in the reference configuration moves to a point $x_i$ in the deformed configuration. The change in shape of infinitesimal volume elements within the solid can then be characterized by the deformation gradient $F_{ij} = \partial x_i / \partial X_j$, in the usual manner. Following the standard framework of finite deformation elasticity and plasticity, we separate the deformation into contributions arising from plastic flow, expansion due to insertion of Li into the network; and an elastic distortion of the lattice, by writing

$$F_{ij} = F^e_{ik} F^c_{kl} F^p_{lj}$$

The total deformation of a representative material element can be visualized as resulting from a sequence of an irreversible plastic deformation, represented by $F^p_{ij}$, followed by a shape change $F^c_{ij}$ induced by intercalating Li at zero stress, followed by a reversible elastic distortion of the network, represented by $F^e_{ij}$. For future reference, we note that $F^c_{ij}$ is a function of the Li concentration $\rho$. The velocity gradient follows as

$$L_{ij} = \frac{\partial \dot{x}_i}{\partial x_j} = \dot{F}_{ik} F^{-1}_{kj} = \dot{F}^e_{ik} F^{e-1}_{kj} + F^e_{in} \dot{F}^c_{nl} F^{c-1}_{lk} F^{e-1}_{kj} + F^e_{in} F^c_{nm} \dot{F}^p_{mr} F^{p-1}_{rl} F^{c-1}_{lk} F^{e-1}_{kj}$$

The three terms in this expression can be interpreted as the velocity gradient due to elastic stretching; due to the insertion of Li; and due to plastic flow, respectively.

*2.2 The Cathode*

The behavior of the cathode is very similar to that of the anode, and so needs only a brief discussion. The cathode is a network solid with molar density $\rho_{0C}$ per unit undeformed volume of lattice sites, which occupies a volume $V_{0C}$ in its initial, stress free configuration. The surface of the cathode is divided into a portion $A_{0Cext}$, which is subjected to external mechanical and electrical loading, together with a portion $A_{0Ce}$ that is adjacent to the electrolyte. The area $A_{0Cext}$ is subjected to traction $t_i$. In addition, the external circuit connecting the anode and cathode leads to a flux of electrons across $A_{0Cext}$. We denote the electron flux density per unit undeformed area flowing out of the cathode across $A_{0Cext}$ and $A_{0Ce}$ by $i_{Cext}$ and $i_{Ce}$, respectively.

In a half-cell, the cathode is solid Li, and Li is removed directly from the cathode at the electrolyte interface. Consequently, the molar density of Li per unit reference volume inside the cathode is constant. The flow of electrons through the cathode satisfies the conservation law (4), as well as boundary conditions of the form (3), with subscripts denoting the anode replaced by those signifying the cathode. The cathode may also change its shape as a result of mechanical loading. For simplicity, we assume that the cathode remains elastic, and has a fixed composition. Its deformation can be completely characterized by the deformation gradient $F_{ij} = \partial x_i / \partial X_j$.

*2.3 The Electrolyte*

Modeling the behavior of battery electrolytes is a major objective of electrochemistry and many sophisticated treatments exist – see for example Guyer *et al* (2004) or Biesheuvel *et al* (2009). Our focus here is on the mechanical processes occurring in the electrodes, and the coupling between mechanics and electrochemistry in the electrode/electrolyte interfaces. Accordingly, we have adopted the simplest possible description of an electrolyte that can provide a thermodynamically consistent derivation of the field equations. Its main purpose is to ensure that charge can be properly conserved in the system. Models somewhat similar to ours are sometimes used to model solid electrolytes with fixed background charge, but a more sophisticated treatment would be needed for a proper description of Li-ion batteries, which contain at least two mobile charged species in the electrolyte.

Accordingly, the electrolyte will be idealized as an inviscid, incompressible fluid, containing a molar density $\rho^+$ of Li$^+$. Although the electrolyte is a fluid, its deformation can be described as though it were an amorphous network solid, which occupies a reference volume $V_{0F}$, and contains a molar density $\rho_{0F}$ of network lattice sites. The Li$^+$ will be treated as a substitutional solute which diffuses through this network. Conservation of mass requires that

$$\dot{\rho}^+ + \frac{\partial j_j^+}{\partial X_j} = 0 \tag{6}$$

within the fluid. The electrolyte contacts the electrodes at the interfaces $A_{Ae}$ and $A_{Ce}$. Here, the molar fluxes of Li-ions $j_A^+, j_C^+$ per unit reference area flow into the electrolyte from the reaction layers at the anode and cathode, respectively. This leads to boundary conditions

$$j_i^+ N_i = j_A^+ \quad X_i \in A_{0Ae} \qquad j_i^+ N_i = j_C^+ \quad X_i \in A_{0Ce} \qquad (7)$$

at the electrode interfaces, where $N_i$ denotes the components of a unit vector normal to the electrode surfaces, pointing into the electrolyte.

The deformation of the electrolyte can be described as though it were a solid. A generic lattice point with position $X_i$ in the reference configuration moves to a point $x_i$ in the deformed configuration. The change in shape of infinitesimal volume elements within the solid can then be characterized by the deformation gradient $F_{ij} = \partial x_i / \partial X_j$. Since the fluid is incompressible, the deformation gradient must satisfy $J=\det(\mathbf{F})=1$ inside the electrolyte, or equivalently $\dot{F}_{ij} F_{ji}^{-1} = 0$

*2.4 Anode/electrolyte interface reaction*

The 'Stern layer' that forms adjacent to the anode/electrolyte interface is idealized as a thin film between the solid and the fluid, as shown in Fig 1. The layer has some small (atomic scale) thickness $a$ in the reference configuration. The interfacial region develops a double-layer of charge, which is produced by a molar concentration $Q$ of electrons per unit undeformed area on the surface of the anode, and a molar concentration $R^+$ of $Li^+$ per unit area on the fluid surface. The region between these two layers is free of charge.

The electrons and ions in the double-layer may combine to form a Li atom, through the reaction $Li^+ + e^- \rightarrow Li$. We denote the number of moles of Li atoms formed by this reaction per unit reference area by $\dot{P}_A$. A molar concentration $R$ per unit reference area of neutral Li atoms is present in the interfacial region as a result of this reaction.

The interfacial region may exchange Li atoms and electrons with the anode, and may also exchange $Li^+$ with the electrolyte. We denote the molar flux of Li per unit undeformed area flowing from the anode to the Stern layer by $j_A$; and the flux of electrons flowing from the anode to the Stern layer by $i_A$. The interface injects a molar flux of ions $j_A^+$ into the electrolyte. Mass conservation implies that

$$\dot{R}^+ + j^+ + \dot{P}_A = 0 \quad \dot{R} - j_A - \dot{P}_A = 0 \quad \dot{Q} - i_{Ae} + \dot{P}_A = 0 \qquad (8)$$

Finally, we must characterize the deformation of the interfacial region. We assume that the deformation inside the layer can be described by a uniform deformation gradient $F_{ij}^*$, which is independent of distance from the anode surface, and which satisfies $\det(F_{ij}^*) = 1$ (i.e. the layer is incompressible). We assume that no volume changes occur when $Li^+$ combine with electrons. In addition, we assume that, when a Li$^+$ ion leaves the Stern layer, it is replaced by a neutrally charged atom with identical volume from the electrolyte. Consequently, volume changes in the fluid and Stern layer occur only when a Li atom leaves the anode and is incorporated into the interfacial region between the electrolyte and anode. Since we have assumed that the Stern layer is incompressible, and has fixed dimensions in the reference configuration, adding a Li atom to the interfacial region must be accommodated by ejecting an equal volume into the electrolyte. The velocities of two material particles in the anode and fluid are therefore related by

$$\dot{x}_i^A + \dot{F}_{ij}^* a N_j = \dot{x}_i^F - \frac{\partial x_i^F}{\partial X_j} N_j \Omega N_a j_A \qquad (9)$$

where $N_j$ is the outward normal to the un-deformed anode, $\Omega$ is the volume of an Li atom, and $N_a$ is the Avogadro number. Here, the two terms on the right hand side represent the velocity of the electrolyte surface that contacts the Stern layer. The first of these describes flow in the fluid; the second accounts for the motion of the surface of the fluid as additional Li atoms leave the anode and are incorporated into the fluid.

We note that descriptions of the double layer in the literature differ widely, see *e.g.* Biesheuvel *et al* (2009) for a review. Our description is similar to that used by Biesheuvel *et al.* (2009), except that we allow singular distributions of charge to develop at both surfaces of the double layer, and use the thickness of the double layer as a constitutive parameter. In contrast, Biesheuvel *et al.* (2009) calculate the double-layer dimensions that lead to a continuous distribution of electric field on one side of the double-layer. The detailed description of the double layer is not the focus of our paper – we will simply adopt our idealization as a working hypothesis and derive appropriate constitutive equations for this model.

*2.5 Cathode/electrolyte interface reaction*

A similar reaction layer is present at the interface between the cathode and electrolyte. Here, Li atoms detach from the cathode surface and are injected into the Stern layer at a rate $j_C$, while a flux of electrons $i_C$ flows from the cathode into the reaction layer. In the reaction layer, Li and Li$^+$ react with electrons Li$^+ + e^- \to$ Li. A flux of Li$^+$, $j^+$, is injected into the electrolyte from the reaction layer.

Mass conservation in the electrochemical reaction requires that

$$\dot{R}^+ + j^+ + \dot{P}_C = 0 \qquad \dot{R} - j_C - \dot{P}_C = 0 \qquad \dot{Q} - i_{Ce} + \dot{P}_C = 0 \qquad (10)$$

As at the anode, we assume that when an Li atom is injected into the Stern layer, a second atom with identical volume is ejected into the fluid electrolyte. Displacement continuity at the electrolyte/cathode interface then requires that

$$\dot{x}_i^C + \dot{F}_{ij}^* a N_j - \frac{\partial x_i^C}{\partial X_j} N_j \Omega N_a j_C = \dot{x}_i^F - \frac{\partial x_i^F}{\partial X_j} N_j \Omega N_a j_C \qquad (11)$$

where $\Omega$ is the volume of a single Li atom in the un-deformed system, and $N_a$ is the Avogadro number.

*2.5 Electric fields*

The charge on electrons and Li$^+$ in the electrolyte (and their fluxes) induce electromagnetic fields in the system. For simplicity, we neglect magnetic fields and polarization. The electric field $\eta_i$ in the electrolyte is related to the concentration of Li$^+$ by Gauss law

$$\frac{\partial \eta_i}{\partial x_i} - \frac{eN_a\left(\rho^+ - \rho_0^+\right)}{J\varepsilon_0} = 0 \qquad (12)$$

where $e$ is the magnitude of the charge on an electron (a positive number), $N_a$ is the Avogadro number[3], $\rho_0^+$ is the concentration of Li$^+$ in an electrically neutral electrolyte, and $\varepsilon_0$ is the permittivity of free space. Similarly, in the anode and cathode

$$\frac{\partial \eta_i}{\partial x_i} + \frac{eN_a(q - q_0)}{\varepsilon_0 J} = 0 \qquad (13)$$

where $q_0$ is the density of free electrons in the electrically neutral solid. In our treatment, singular distributions of charge can only develop at the electrode/electrolyte interfaces. In addition, we assume that charge is conserved within the system. Consequently, we may simplify calculations by assuming that the electric field outside the system vanishes, which implies that $\eta_i = 0$ at exterior surfaces of the electrodes. In addition, the electric field must be continuous across the electrode/electrolyte interfaces.

It is convenient to re-write (12) and (13) in variational form, which allows us to define measures of electric fields for the undeformed anode and cathode. Let $\phi$ be any admissible distribution of electric potential within the solid, which must be a continuous function of position. Then

$$\int_{V_A}\left(\frac{\partial \eta_i}{\partial x_i} + \frac{eN_a(q-q_0)}{\varepsilon_0 J}\right)\phi dV + \int_{V_C}\left(\frac{\partial \eta_i}{\partial x_i} + \frac{eN_a(q-q_0)}{\varepsilon_0 J}\right)\phi dV$$
$$+ \int_{V_F}\left(\frac{\partial \eta_i}{\partial x_i} - \frac{eN_a(\rho^+ - \rho_0^+)}{\varepsilon_0}\right)\phi dV = 0 \qquad (14)$$

Integrating by parts, using the boundary conditions on electric field at external surfaces and the electrode/electrolyte interfaces, and transforming the volume integrals in the anode and cathode to the reference configuration, this expression can be re-written as

$$\int_{V_{0A}} \left( \varepsilon_0 H_j \frac{\partial \phi}{\partial X_j} - eN_a(q - q_0)\phi \right) dV_0 + \int_{V_{0C}} \left( \varepsilon_0 H_j \frac{\partial \phi}{\partial X_j} - eN_a(q - q_0)\phi \right) dV_0$$

$$+ \int_{V_F} \left( \varepsilon_0 H_j \frac{\partial \phi}{\partial X_j} + eN_a(\rho^+ - \rho_0^+)\phi \right) dV - \int_{A_{Ae}} \varepsilon_0 \left( H_j^A N_j \phi_A - H_j^F N_j \phi_F \right) dA \quad (15)$$

$$- \int_{A_{Ce}} \varepsilon_0 \left( H_j^C N_j \phi_C - H_j^F N_j \phi_F \right) dA = 0$$

where

$$H_j = JF_{ji}^{-1} \eta_i \quad (16)$$

is an electric field defined on the undeformed solid, and where $H_i^A, H_i^F, H_i^C$ denote the electric fields just outside the Stern layers in the anode, electrolyte and cathode, respectively. It is straightforward to show that

$$\frac{\partial H_j}{\partial X_j} + \frac{eN_a(q - q_0)}{\varepsilon_0} = 0 \qquad \frac{\partial H_i}{\partial X_i} - \frac{eN_a\left(\rho^+ - \rho_0^+\right)}{\varepsilon_0} = 0 \quad (17)$$

in both anode, cathode, and electrolyte.

---

[3] The product $eN_a$ is the Faraday constant, and is often denoted by $F$. We do not use this notation here to avoid confusion with the deformation gradient.

Inside the Stern layers, the electric field $H_j^*$ is uniform, and the tangential component of electric field must be continuous across the boundaries between Stern layers and the anode, electrolyte and cathode. The normal component is discontinuous across these boundaries. For example, the fields $H_i^F$, $H_i^A$ in the fluid and anode adjacent to the Stern layer are related to $H_j^*$ by

$$H_i^F N_i - H_i^* N_i - \frac{eN_a(R^+ - R_0^+)}{\varepsilon_0} = 0 \qquad H_i^* N_i - H_i^A N_i + \frac{eN_a(Q - Q_0)}{\varepsilon_0} = 0 \quad (18)$$

where $R_0^+, Q_0^+$ are the molar densities of Li$^+$ and electrons per unit area of uncharged reference surface.

Finally, an expression for the electrostatic energy density in the system will be useful. The total electrostatic energy is related to the electric fields by

$$G^e = \frac{1}{2} \int_V \varepsilon_0 \eta_i \eta_i \, dV \quad (19)$$

Where the integral is taken over the entire volume of the solid, including anode, cathode, and Stern layers. This may be re-written as an integral over the reference configuration as

$$G^e = \frac{1}{2}\int_{V_0} \frac{\varepsilon_0}{J} H_j F_{ij} F_{ik} H_k dV \qquad (20)$$

*2.6 Mechanical equilibrium*

We assume *a priori* that the structure is in static equilibrium. This implies that a Cauchy stress field $\sigma_{ij}$ acts in the system, which must satisfy $\sigma_{ij} n_j = t_i$ on exterior boundaries, and the equilibrium equation

$$\frac{\partial \sigma_{ij}}{\partial x_i} - \frac{eN_a}{J}(q-q_0)\eta_j = 0 \quad x_i \in V_A \cup V_C \qquad \frac{\partial \sigma_{ij}}{\partial x_i} + \frac{eN_a}{J}(\rho^+ - \rho_0^+)\eta_j = 0 \quad x_i \in V_F \qquad (21)$$

in the anode, cathode, and electrolyte. The second term in each equilibrium equation represents the body force associated with the electric charge in the system. It is helpful to re-write the equilibrium equation in variational form. To this end, let $\delta \dot{x}_i$ be any kinematically admissible velocity field in the system, which must be continuous in the electrodes and electrolyte, but which may have discontinuities across the electrode/electrolyte interfaces. Multiply each equation in (21) by $\delta \dot{x}_i$, integrate over the volume of the system, express the electron and ion concentrations in terms of the electric fields by using eqs (12) and (13) to see that

$$\int_{V_A+V_F+V_C} \frac{\partial}{\partial x_i}\left(\sigma_{ij} + \sigma_{ij}^M\right)\delta \dot{x}_i dV = 0 \qquad (22)$$

where

$$\sigma_{ij}^M = \varepsilon_0\left(\eta_i \eta_j - \frac{1}{2}\eta_k \eta_k \delta_{ij}\right) \qquad (23)$$

is a tensor valued representation of the body forces induced by electrostatic interactions, known as the 'Maxwell stress.' Its physical significance can be clarified by noting that

$$\frac{\partial}{\partial x_i}\varepsilon_0\left(\eta_i \eta_j - \frac{1}{2}\eta_k \eta_k \delta_{ij}\right) = \varepsilon_0 \frac{\eta_i}{\partial x_i}\eta_j \qquad (24)$$

where the right hand side can be recognized as the body force acting on the charge distribution in the system. Now, eq. (22) may be integrated by parts, giving

$$\int_{V_A}\left(\sigma_{ij} + \sigma_{ij}^M\right)\delta L_{ij} dV + \int_{V_F}\left(\sigma_{ij} + \sigma_{ij}^M\right)\delta L_{ij} dV + \int_{V_C}\left(\sigma_{ij} + \sigma_{ij}^M\right)\delta L_{ij} dV$$
$$= \int_{A_{Aext}} t_i \delta \dot{x}_i dA + \int_{A_{Cext}} t_i \delta \dot{x}_i dA + \int_{A_{Ae}} (\tilde{\sigma}_{ij}^A \delta \dot{x}_i^A - \tilde{\sigma}_{ij}^F \delta \dot{x}_i^F)n_j dA + \int_{A_{Ce}} (\tilde{\sigma}_{ij}^C \delta \dot{x}_i^C - \tilde{\sigma}_{ij}^F \delta \dot{x}_i^F)n_j dA \qquad (25)$$

where $\delta L_{ij} = \partial \delta \dot{x}_i / \partial x_j$ is the velocity gradient, $\tilde{\sigma}_{ij}^A, \tilde{\sigma}_{ij}^F, \tilde{\sigma}_{ij}^C$ denote the combined mechanical and Maxwell stresses in the solid adjacent to the electrode/electrolyte interfaces in the anode, electrolyte, and cathode, respectively, and $n_j$ is the outward normal to the electrodes. For later calculations, it is convenient to re-write this expression in the form

$$\int_{V_{0A}} \left( \sigma_{ij} + \sigma_{ij}^M \right) \left( \dot{F}_{ik}^e F_{kj}^{e-1} + F_{in}^e \dot{F}_{nl}^c F_{lk}^{c-1} F_{kj}^{e-1} + F_{in}^e F_{nm}^c \dot{F}_{mr}^p F_{rl}^{p-1} F_{lk}^{c-1} F_{kj}^{e-1} \right) J dV$$
$$+ \int_{V_{0F}} \left( \sigma_{ij} + \sigma_{ij}^M \right) \dot{F}_{ik} F_{kj}^{-1} J dV + \int_{V_{0C}} \left( \sigma_{ij} + \sigma_{ij}^M \right) \dot{F}_{ik} F_{kj}^{-1} J dV \quad (26)$$
$$= \int_{A_{Aext}} t_i \dot{x}_i dA + \int_{A_{Cext}} t_i \dot{x}_i dA + \int_{A_{Ae}} (\tilde{\sigma}_{ij}^A \delta \dot{x}_i^A - \tilde{\sigma}_{ij}^F \delta \dot{x}_i^F) n_j dA + \int_{A_{Ce}} (\tilde{\sigma}_{ij}^C \delta \dot{x}_i^C - \tilde{\sigma}_{ij}^F \delta \dot{x}_i^F) n_j dA$$

where $J = \det(\mathbf{F})$.

## 3. Thermodynamics

A standard thermodynamic argument provides the generalized thermodynamic forces conjugate to the kinematic quantities identified in the preceding section. The anode, electrolyte, cathode, and reaction layers together form a thermodynamic system. The total free energy of this system is

$$G = \int_{V_{0A}} \psi^A(\mathbf{F}^e, \mathbf{F}^p, \rho, q, H_i) dV + \int_{V_{0F}} \psi^F(\mathbf{F}, \rho^+, H_i) dV + \int_{V_{0C}} \psi^C(\mathbf{F}, q, H_i) dV$$
$$+ \int_{A_{0Ce}} \Psi^A(\mathbf{F}^*, R^+, Q, R, H_i^*) dA + \int_{A_{0Ce}} \Psi^C(\mathbf{F}^*, R^+, Q, R, H_i^*) dA - \int_{A_{Aext}} t_i x_i dA - \int_{A_{Cext}} t_i x_i dA \quad (27)$$

Here, $\psi^A, \psi^F, \psi^C$ denote the free energy per unit undeformed volume in the anode, electrolyte and cathode, respectively, while $\Psi^A, \Psi^C$ are the free energies of the Stern layer at the anode and cathode interfaces, respectively. The terms involving tractions represent the potential energy of external loads. The energy associated with electrostatic fields is included through the dependence of the free energies on the electric fields $H_i$.

The second law of thermodynamics requires that the rate of change of free energy in this system must be negative or zero for any compatible variation in displacement, charge, electric field, and Li (or Li-ion) concentration field in this system. We proceed to calculate the rate of change of free energy, introducing Lagrange multipliers to enforce the conservation laws listed in the preceding section. Specifically, conservation of electrons will be enforced by introducing the electrochemical potential $\mu_{Aq}, \mu_{Cq}$ of the electrons in the anode and cathode; conservation of Li in the anode will be enforced by introducing the chemical potential $\mu_A$ of interstitial Li; conservation of Li$^+$ in the electrolyte is enforced by means of the electrochemical potential $\mu^+$ of Li$^+$ in the electrolyte, while volume preservation in the electrolyte is enforced by the electrolyte pressure $\pi$. Similarly, we introduce chemical potentials $M_{AR}, M_{AQ}$ to enforce conservation of mass of Li, Li$^+$ and electrons in the Stern layer at the anode, with a corresponding set at the cathode. Volume preservation in the Stern layer is enforced by a Lagrange multiplier $\pi_{ij}^*$. Finally, conservation of electrons flowing through the external circuit is enforced by means of a Lagrange multiplier $\Lambda$.

A lengthy calculation, which is described in detail in the Appendix, shows that the rate of change of free energy can be separated into contributions from the anode, fluid electrolyte, cathode, and the Stern layers at the anode and cathode interfaces

$$\frac{\partial G}{\partial t} = \frac{\partial G^A}{\partial t} + \frac{\partial G^C}{\partial t} + \frac{\partial G^F}{\partial t} + \frac{\partial G^{Ae}}{\partial t} + \frac{\partial G^{Ce}}{\partial t} + \frac{\partial G^{ext}}{\partial t} \qquad (28)$$

where

$$\begin{aligned}
\frac{\partial G^A}{\partial t} &= \int_{V_{0A}} \left( \frac{\partial \psi^A}{\partial F_{ij}^e} - J(\sigma_{ik} + \sigma_{ik}^M) F_{jk}^{e-1} \right) \dot{F}_{ij}^e dV + \int_{V_{0A}} \left( \frac{\partial \psi^A}{\partial F_{ij}^p} - J(\sigma_{pq} + \sigma_{pq}^M) F_{pn}^e F_{ni}^c F_{jl}^{p-1} F_{lk}^{c-1} F_{kq}^{e-1} \right) \dot{F}_{ij}^p dV \\
&+ \int_{V_{0A}} \left( \frac{\partial \psi^A}{\partial \rho} - J(\sigma_{ij} + \sigma_{ij}^M) F_{in}^e \frac{\partial F_{nl}^c}{\partial \rho} F_{lk}^{c-1} F_{kj}^{e-1} - \mu_A \right) \dot{\rho} dV + \int_{V_{0A}} \left( \frac{\partial \psi^A}{\partial q} - eN_a \phi - \mu_{Aq} \right) \dot{q} dV \\
&+ \int_{V_{0A}} \left( \frac{\partial \psi^A}{\partial H_j} + \frac{\partial \phi}{\partial X_j} \right) \dot{H}_j dV + \int_{V_{0A}} \frac{\partial \mu_{qA}}{\partial X_j} i_j dV + \int_{V_{0A}} \frac{\partial \mu_A}{\partial X_j} j_j dV
\end{aligned} \qquad (29)$$

is the rate of change of free energy associated with processes that take place in the anode;

$$\begin{aligned}
\frac{\partial G^C}{\partial t} &= + \int_{V_{0C}} \left( \frac{\partial \psi^C}{\partial F_{ij}} - J(\sigma_{ik} + \sigma_{ik}^M) F_{jk}^{-1} \right) \dot{F}_{ij} dV + \int_{V_{0C}} \left( \frac{\partial \psi^C}{\partial q} - eN_a \phi - \mu_{Cq} \right) \dot{q} dV \\
&+ \int_{V_{0C}} \left( \frac{\partial \psi^C}{\partial H_j} + \frac{\partial \phi}{\partial X_j} \right) \dot{H}_j dV + \int_{V_{0C}} \frac{\partial \mu_{qC}}{\partial X_j} i_j dV
\end{aligned} \qquad (30)$$

is the rate of change of free energy in the cathode,

$$\begin{aligned}
\frac{\partial G^F}{\partial t} &= \int_{V_{0F}} \left( \frac{\partial \psi^F}{\partial \rho^+} + eN_a \phi - \mu^+ \right) \dot{\rho}^+ dV + \int_{V_{0F}} \left( \frac{\partial \psi^F}{\partial F_{ij}} - J(\sigma_{ik} + \sigma_{ik}^M) F_{jk}^{-1} - J\pi \delta_{ik} F_{jk}^{-1} \right) \dot{F}_{ij} \\
&+ \int_{V_{0F}} \left( \frac{\partial \psi^F}{\partial H_j} + \frac{\partial \phi}{\partial X_j} \right) \dot{H}_j dV + \int_{V_{0F}} \frac{\partial \mu^+}{\partial X_j} j_j^+ dV
\end{aligned} \qquad (31)$$

is the rate of change of free energy in the electrolyte;

$$\begin{aligned}
\frac{\partial G^{AE}}{\partial t} &= \int_{A_{0Ae}} (\tilde{\sigma}_{ij}^A - \tilde{\sigma}_{ij}^F) n_j \dot{x}_i^A dA + \int_{A_{0Ae}} \left( \frac{\partial \Psi^A}{\partial F_{ij}^*} - aJ \left( \pi_{ik}^* F_{jk}^{*-1} + \tilde{\sigma}_{ik}^F F_{lk}^{*-1} N_l N_j \right) \right) \dot{F}_{ij}^* dA \\
&+ \int_{A_{0Ae}} (\frac{\partial \Psi^A}{\partial H_i^*} - \varepsilon_0 N_i \phi_A + \varepsilon_0 N_i \phi_F) \dot{H}_i^* dA + \int_{A_{0Ae}} (\frac{\partial \Psi^A}{\partial R} - M_{AR}) \dot{R} dA \\
&+ \int_{A_{0Ae}} (\frac{\partial \Psi^A}{\partial R^+} + e N_a \phi_F - M_{A+}) \dot{R}^+ dA + \int_{A_{0Ae}} (\frac{\partial \Psi^A}{\partial Q} - e N_a \phi_A - M_{AQ}) \dot{Q} dA \\
&- \int_{A_{0Ae}} (M_{A+} + N_k F_{ik} J \tilde{\sigma}_{ij}^F F_{lj}^{-1} N_l \Omega N_a - \mu^+) j_{Ae}^+ dA - \int_{A_{0Ae}} (\mu_A - M_{AR}) j_A dA \\
&- \int_{A_{0Ae}} (\mu_{Ae} - M_{AQ}) i_{Ae} dA - \int_{A_{0Ae}} (M_{A+} + M_{AQ} - M_{AR}) \dot{P}_A
\end{aligned} \quad (32)$$

is the rate of change of free energy in the Stern layer at the anode, and

$$\begin{aligned}
\frac{\partial G^{CE}}{\partial t} &= \int_{A_{0Ae}} (\tilde{\sigma}_{ij}^C - \tilde{\sigma}_{ij}^C) n_j \dot{x}_i^C dA + \int_{A_{0Ae}} \left( \frac{\partial \Psi^C}{\partial F_{ij}^*} - aJ \left( \pi_{ik}^* F_{jk}^{*-1} + \tilde{\sigma}_{ik}^F F_{lk}^{*-1} N_l N_j \right) \right) \dot{F}_{ij}^* dA \\
&+ \int_{A_{0Ce}} (\frac{\partial \Psi^C}{\partial H_i^*} - \varepsilon_0 N_i \phi_C + \varepsilon_0 N_i \phi_F) \dot{H}_i^* dA + \int_{A_{0Ce}} (\frac{\partial \Psi^C}{\partial R} - M_{CR}) \dot{R} dA \\
&+ \int_{A_{0Ce}} (\frac{\partial \Psi^C}{\partial R^+} + e N_a \phi_F - M_{C+}) \dot{R}^+ dA + \int_{A_{0Ce}} (\frac{\partial \Psi^C}{\partial Q} - e N_a \phi_C - M_{CQ}) \dot{Q} dA \\
&- \int_{A_{0Ce}} (M_{C+} + F_{ik}^F N_k J \tilde{\sigma}_{ij}^F F_{lj}^{-1} N_l \Omega N_a - \mu^+) j_C^+ dA \\
&- \int_{A_{0Ce}} ((\psi^C - \Psi^C \kappa - F_{ik}^C N_k \tilde{\sigma}_{ij}^C JF_{lj}^{C-1} N_l) \Omega N_a - M_{CR}) j_C dA \\
&- \int_{A_{0Ce}} (\mu_{Cq} - M_{CQ}) i_{Ae} dA - \int_{A_{0Ce}} (M_{C+} + M_{CQ} - M_{CR}) \dot{P}_C
\end{aligned} \quad (33)$$

is the rate of change of free energy in the Stern layer between the cathode and electrolyte. In this expression, $F_{ij}^F, F_{ij}^C$ denote the deformation gradient in the electrolyte and cathode, respectively. Finally

$$\frac{\partial G^{ext}}{\partial t} = -\left( \int_{A_{0Aext}} (\Lambda + \mu_{Aq}) i_{Aext} dA + \int_{A_{0Cext}} (\Lambda + \mu_{Cq}) i_{Cext} dA \right) \quad (34)$$

is the rate of change of free energy resulting from current flow through the external circuit.

Each integral in equations (29)-(33) consists of a product between a flux and a generalized thermodynamic force conjugate to the flux. Constitutive equations must relate these two quantities. Any constitutive law is permissible, provided that the rate of change of free energy is negative or zero for all possible fluxes. In the next section, we list constitutive equations that meet these requirements, and also reduce to standard forms used to model the mechanical response of solids, and electrochemical reactions in the absence of stress.

## 4. Constitutive Equations

If a suitable form is selected for the free energy density in each part of the cell, the expressions for the rate of change of free energy listed in the preceding section generate a series of boundary conditions, field equations, and constitutive equations for each region of the solid. We proceed to work through the various components of the electrochemical cell in turn.

*4.1 The Anode and cathode*

To approximate the free energy of the anode, we make two simplifying assumptions: (i) The stress-free strain due to Li insertion consists of a volume change, without shape change; (ii) Although the volumetric and plastic strains may be large, the reversible elastic distortion of the lattice is a small deformation. With this in mind, we assume that the deformation gradient representing expansion of the anode due to Li insertion is

$$F_{ij}^c = \beta(c)^{1/3} \delta_{ij} \qquad c = \rho / \rho_{0A} \tag{35}$$

where $c$ is the molar concentration of Li in the Si, and $\beta(c)$ represents the volume expansion ratio of the stress-free anode when a molar concentration $c$ of Li intercalates into the solid. We also define the elastic Lagrange strain tensor

$$E_{ij}^e = \left( F_{ki}^e F_{kj}^e - \delta_{ij} \right) / 2 \tag{36}$$

and the plastic velocity gradient, stretch rate and spin tensors

$$L_{pq}^p = F_{pn}^e F_{ni}^c \dot{F}_{ij}^p F_{jl}^{p-1} F_{lk}^{c-1} F_{kq}^{e-1} \qquad D_{ij}^p = (L_{ij}^p + L_{ji}^p)/2 \qquad W_{ij}^p = (L_{ij}^p - L_{ji}^p)/2 \tag{37}$$

We then select a free energy density function for the anode given by

$$\psi^A(\mathbf{F}^e, \mathbf{F}^p, \rho, q, H_i) = \rho_{0A} \psi_0^A(c,q) + C_{ijkl}(c) E_{ij}^e E_{kl}^e / 2 + \frac{\varepsilon_0}{2J} H_j F_{ij} F_{ik} H_k \tag{38}$$

where $\psi_0^A(c,q)$ is the free energy per mol of reference lattice for a stress and charge free anode containing a molar concentration $c$ of interstitial Li; and $C_{ijkl}$ is the tensor of elastic moduli, which depends on the Li concentration. The first term in eq. (38) represents chemical energy; the second term is the strain energy density; and the third term is the electrostatic energy density.

With this choice of free energy, the thermodynamic relation in eq. (29) yields expressions relating the stress to the deformation; the chemical potential to stress; and kinetic relations for diffusion of electrons and Li through the anode. For example, the first integral in (29) must be negative or zero for all possible $\dot{F}^e_{ij}$, which shows that the Cauchy stress is related to elastic Lagrange strain by

$$J(\sigma_{ik} + \sigma^M_{ik})F^{e-1}_{jk} = \frac{\partial \psi^A}{\partial F^e_{ij}} = C_{njkl}E^e_{kl}F^e_{in} + J\sigma^M_{ik}F^{e-1}_{jk} \Rightarrow J\sigma_{ij} \approx C_{ijkl}E^e_{kl} \quad (39)$$

Here we have made use of the assumption that elastic distortions are small, and we have noted that

$$\frac{\partial}{\partial F^e_{pq}}\left(\frac{\varepsilon_0}{2J}H_jF_{ij}F_{ik}H_k\right) = \frac{\varepsilon_0}{J}\left(F_{pk}H_kF_{jl}H_l - \frac{1}{2}F_{nk}H_kF_{nl}H_l\delta_{pj}\right)F^{e-1}_{qj}$$
$$= J\varepsilon_0\left(\eta_i\eta_j - \frac{1}{2}\eta_k\eta_k\delta_{ij}\right)F^{e-1}_{qj} = J\sigma^M_{pj}F^{e-1}_{qj} \quad (40)$$

The elastic moduli in eq. (39) must be interpreted carefully. By definition, they relate the *Kirchhoff* stress in the solid to the elastic Lagrange strain. The Jacobian of the deformation gradient $J$ includes the effects of volume expansion due to lithiation, and may greatly exceed unity. For example, a fully lithiated Si electrode has $J \approx 3$. If the elastic constants $C_{ijkl}$ are independent of Li concentration, eq. (39) predicts that the slope of the Cauchy-stress –v- elastic strain curve will be reduced by a factor of 3 after the electrode is lithiated. A somewhat smaller reduction in modulus has been predicted by first principles studies by Shenoy *et al.* (2010) and measured experimentally by Sethuraman et al. (2010d), suggesting that in practice the moduli $C_{ijkl}$ increase slightly with Li concentration.

We next derive constitutive equations for the plastic strain rate. The second integral in (29), together with the choice of free energy in (38) and a similar result to eq. (40) (but with $F^e_{pq}$ replaced by $F^p_{pq}$) shows that the constitutive law for plastic strains must satisfy

$$J\sigma_{pq}F^e_{pn}F^c_{ni}\dot{F}^p_{ij}F^{p-1}_{jl}F^{c-1}_{lk}F^{e-1}_{kq} \geq 0 \Rightarrow J\sigma_{ij}D^p_{ij} \geq 0 \quad (41)$$

for all possible stress states and plastic strain rates. This condition can be met by a standard viscoplastic constitutive equation (Gurtin et al, 2010). For simplicity, we take the anode to be isotropic, and assume that volumetric plastic strains vanish. Under these conditions, the plastic strain rate can be computed from a viscoplastic flow potential $\Upsilon(\tau_e)$, where $\tau_e$ is an effective Kirchhoff stress defined as

$$\tau_e = \sqrt{3\tau^D_{ij}\tau^D_{ij}/2} \qquad \tau^D_{ij} = J\sigma_{ij} - J\sigma_{kk}\delta_{ij}/3 \quad (42)$$

The plastic stretch rate $D^p_{ij}$ is then related to the flow potential by

$$D_{ij}^p = \frac{\partial \Upsilon}{\partial \tau_{ij}} \tag{43}$$

The plastic flow potential may be any monotonically increasing function of $\tau_e$. Here, we will adopt a power-law of the form

$$\Upsilon(\tau_e) = \begin{cases} \dfrac{\sigma_0 \dot{d}_0}{m+1}\left(\dfrac{\tau_e}{\sigma_0}-1\right)^{m+1} & \tau_e > \sigma_0 \\ 0 & \tau_e < \sigma_0 \end{cases} \tag{44}$$

where $m, \dot{d}_0, \sigma_0$ are material constants (which may depend on Li concentration).

A constitutive law must also specify the plastic spin $W_{ij}^p$. Here, we simply take $W_{ij}^p = 0$, following the usual assumption for isotropic viscoplastic solids.

The third integral in (29), together with the choice of free energy in (38) defines the chemical potential of Li in the anode. By assuming that elastic strains are small, and that plastic strains are volume preserving, the result may be simplified to

$$\mu_A = \frac{\partial \psi_0^A}{\partial c} + \frac{1}{2\rho_{0A}}\frac{\partial C_{ijkl}}{\partial c} E_{ij}^e E_{kl}^e - \frac{\sigma_{kk}}{3\rho_{0A}}\frac{\partial \beta}{\partial c} \tag{45}$$

This expression may also be written in terms of stress, by introducing the compliance tensor $S_{ijkl}(c)$ (the inverse of the stiffness tensor). Straightforward algebraic manipulations yield

$$\mu_A = \frac{\partial \psi_0^A}{\partial c} - \frac{\beta^2(c)}{2\rho_{0A}}\frac{\partial S_{ijkl}}{\partial c}\sigma_{ij}\sigma_{kl} - \frac{\sigma_{kk}}{3\rho_{0A}}\frac{\partial \beta}{\partial c} \tag{46}$$

where we have assumed $J \approx \beta$. For infinitesimal deformations, this result reduces to the familiar Larché-Cahn chemical potential of an interstitial solute atom. In practice, the last term in eq. (46) is significantly greater than the term involving the compliance [Sethuraman *et al.* (2010c)]. If this is the case, the small strain chemical potential can be used in finite strain applications, provided that Cauchy stress is used as the stress measure.

The fourth integral in (29) defines the electrochemical potential of electrons in the anode as

$$\mu_{qA} = \rho_{0A}\frac{\partial \psi_0^A}{\partial q} - eN_a\phi \tag{47}$$

The fifth integral in (29) yields the governing equation for the electric potential in the anode

$$\frac{\partial \phi}{\partial X_j} = -\frac{\partial \psi^A}{\partial H_j} = -\frac{\varepsilon_0}{J} F_{ij} F_{ik} H_k \quad \Rightarrow \quad \frac{\partial \phi}{\partial x_j} = -\varepsilon_0 \eta_j \tag{48}$$

Finally, the constitutive equations relating transport of electrons and Li through the anode must ensure that sixth and seventh integrals in (29) are negative or zero. This can be achieved by adopting standard linear relations between flux and potential gradient:

$$i_i = -qD_q \frac{\partial \mu_{qA}}{\partial X_j} \qquad j_i = -\rho D \frac{\partial \mu_A}{\partial X_j} \tag{49}$$

where $D_q$ and $D$ represent the mobility of electrons and Li atoms in the anode, respectively.

A similar approach yields constitutive equations for the cathode, but the calculations are straightforward and will not be described in detail here. The deformation in the cathode is governed by the usual field equations of linear elasticity. In addition, electron flow through the cathode is governed by equations similar to (47) and the first of (49). No mass flows through the interior of the cathode.

*4.2 The electrolyte*

We take the free energy of the electrolyte to be

$$\psi^F(F_{ij}, \rho^+, H_j) = \psi_0^F(\rho^+) + \frac{\varepsilon_0}{2J} H_j F_{ij} F_{ik} H_k \tag{50}$$

The first integral in eq. (31) then defines the electrochemical potential of Li$^+$ in the electrolyte as

$$\mu^+ = \frac{\partial \psi_0^F}{\partial \rho^+} + eN_a \phi \tag{51}$$

The stress field in the electrolyte follows from the second integral in (31), which shows that $\sigma_{ij} = -\pi \delta_{ij}$.

Finally, diffusion of Li$^+$ through the electrolyte is governed by

$$j_i^+ = -\rho^+ D_+ \frac{\partial \mu^+}{\partial X_j} \tag{52}$$

where $D_+$ is the mobility of Li$^+$ in the electrolyte.

*4.3 The electrode/electrolyte interfaces*

Equation (32) provides a series of boundary conditions for stress and electrochemical potentials at the interface between the Stern layer, the anode and the fluid electrolyte. It also defines the driving force for the electrochemical reaction within the Stern layer.

For example, the first integral in (32) gives the boundary conditions relating stress in the electrolyte and in the anode. The integral must vanish for all possible values of velocity in the anode $\dot{x}_i^A$, which shows that

$$\tilde{\sigma}_{ij}^A n_j - \tilde{\sigma}_{ij}^F n_j = 0 \Rightarrow \sigma_{ij}^A n_j - \sigma_{ij}^F n_j = -\left(\sigma_{ij}^{MA} n_j - \sigma_{ij}^{MF} n_j\right) \tag{53}$$

where $\sigma_{ij}^M$ is the Maxwell stress (representing the body forces due to electric charge), defined in eq. (23), and superscripts $A$ and $F$ denote quantities in the anode and electrolyte, respectively. If the net charge in the Stern layer is zero, the electric field is continuous across the Stern layer, so that the right hand side of (53) is zero.

The second integral in eq. (32) yield equations defining the stress $\pi_{ij}^*$ in the Stern layer. This stress acts so as to ensure that the volume of the Stern layer remains constant, but in our model does not influence the electrochemical reaction in any way, so it need not be evaluated. The third integral relates the electric field $H_i^*$ in the Stern layer to the potential difference across the interface

$$\phi_A - \phi_F = F_{ij}^* N_j F_{ik}^* H_k^* a \tag{54}$$

The fifth, sixth and seventh integrals in (32) define the chemical potentials $M_{AR}, M_{A+}, M_{AQ}$ of Li atoms, Li$^+$, and electrons in the Stern layer

$$M_{AR} = \frac{\partial \Psi^A}{\partial R} \qquad M_{A+} = \frac{\partial \Psi^A}{\partial R^+} + eN_a\phi_F \qquad M_{AQ} = \frac{\partial \Psi^A}{\partial Q} - eN_a\phi_A \tag{55}$$

The next three integrals give equilibrium conditions between the Stern layer and the surrounding electrolyte and anode

$$M_{A+} = \mu^+ - N_k F_{ik} J \tilde{\sigma}_{ij}^F F_{lj}^{-1} N_l \Omega N_a \qquad M_{AQ} = \mu_{Ae} \qquad M_{AR} = \mu_A \tag{56}$$

The expression for $M_{A+}$ can be simplified by noting that the stress in the electrolyte is $\sigma_{ij}^F = -\pi\delta_{ij} + \sigma_{ij}^{MF}$, where $\pi$ is the pressure, and $\sigma_{ij}^{MF}$ is the electromagnetic Maxwell stress, defined in eq. (23). This yields

$$M_{A+} = \mu^+ - N_k F_{ik} J \sigma_{ij}^{MF} F_{lj}^{-1} N_l \Omega N_a + J\Omega N_a \pi \tag{57}$$

Similar relations hold at the cathode. Expressions for stress boundary conditions, potential difference and chemical potential are identical to eqns. (53)-(55), (with superscript or subscript $A$ replaced by $C$). The equilibrium of chemical potential between the Stern layer and the adjacent cathode and electrolyte at the cathode gives

$$\begin{aligned} M_{C+} &= \mu^+ - F_{ik}^F N_k J \tilde{\sigma}_{ij}^F F_{lj}^{F-1} N_l \Omega N_a \\ M_{CR} &= (\psi^C - \Psi^C \kappa - F_{ik}^C N_k \tilde{\sigma}_{ij}^C J F_{lj}^{C-1} N_l) \Omega N_a \\ M_{CQ} &= \mu_{Cq} \end{aligned} \quad (58)$$

Here, $F_{ij}^F$ is the deformation gradient in the electrolyte just outside the reaction layer; $F_{ij}^C$ is the corresponding deformation gradient in the cathode; $N_i$ is a unit vector normal to the surface of the undeformed cathode; $\tilde{\sigma}_{ij}^F = \sigma_{ij}^{MF} - \pi \delta_{ij}$, where $\sigma_{ij}^{MF}$ and $\pi$ are the Maxwell stress and pressure in the electrolyte; $\tilde{\sigma}_{ij}^C = \sigma_{ij}^{MC} + \sigma_{ij}^C$, where $\sigma_{ij}^{MF}$ and $\sigma_{ij}^C$ are the Maxwell stress and Cauchy stress in the cathode; $\psi^C$ is the free energy per unit reference volume of the Li cathode; $\Psi^C$ is the free energy per unit area of the reaction layer, and $\kappa$ is the curvature of the surface of the undeformed cathode.

*4.6 Electrochemical reactions*

The last integrals in (32) and (33) quantify the rate of change of free energy arising from the electrochemical reactions that take place at the electrolyte/electrode interfaces. We relate these to the reaction rate using a simple phenomenological constitutive equation (based loosely on transition state theory), which is widely used in describing electrochemical reactions.

A brief summary of the assumptions underlying this model is helpful [Hamann *et al.* (1998)]. We consider the reaction $Li^+ + e^- \to Li$ at the anode, which is assumed to take place by forming an activated complex $(Lie)^{\pm}$, which is in equilibrium with the reactants, but which may also decompose to the reaction product Li. We denote the concentration of the activated complex by $C^{\pm}$. The rate of reaction is then $\nu C^{\pm}$ where $\nu$ is a characteristic attempt frequency. Since the reactants are in equilibrium with the activated complex, their electrochemical potentials are related by

$$M^{\pm} = M_{A+} + M_{AQ} \quad (59)$$

where $M^{\pm}$ is the electrochemical potential of the activated compound, and $M_{A+}$ and $M_{AQ}$ are the electrochemical potentials of the reactants, defined in eq. (55). The chemical potentials of the reactants can be expressed in terms of their concentrations as

$$\frac{\partial \Psi^A}{\partial R^+} = \bar{R}T \log\left(\frac{\Gamma^+ R^+}{R_0^+}\right) \qquad \frac{\partial \Psi^A}{\partial Q} = \bar{R}T \log\left(\frac{\Gamma_Q Q}{Q_0}\right) \quad (60)$$

where $\bar{R}$ is the gas constant, $T$ is temperature, $\Gamma^+$ and $\Gamma_Q$ are dimensionless activity coefficients (which may be functions of concentration) and $R_0^+$ and $Q_0$ are reference concentrations of Li$^+$ and free electrons, respectively. We take the electrochemical potential of the activated complex to be

$$M^\pm = \bar{R}T \log \frac{\Gamma^\pm C^\pm}{C_0^\pm} + \alpha e N_a (\phi_F - \phi_A) \tag{61}$$

where $0 < \alpha < 1$ is a phenomenological 'symmetry factor' that quantifies the influence of the potential difference across the reaction layer on the electrochemical potential of the activated compound. With this choice, eq. (59) can be solved for the concentration of the activated compound, which then gives the rate of reaction as

$$\dot{P}_+ = K_+ Q R^+ \exp[(1-\alpha)(\phi_F - \phi_A) e N_a / \bar{R}T] \tag{62}$$

Here, $K_+$ is a characteristic reaction rate, which can be expressed as a function of the attempt frequency $\nu$ and the activities and reference concentrations of the various species involved in the reaction. An exactly similar procedure can be used to obtain a similar expression for the rate of reverse reaction Li $\to$ Li$^+$ + $e^-$. The net forward reaction then follows as

$$\dot{P}_A = \dot{P}_+ - \dot{P}_- = K_+ Q R^+ \exp[(1-\alpha)(\phi_F - \phi_A) e N_a / \bar{R}T] - K_- R \exp[-\alpha(\phi_F - \phi_A) e N_a / \bar{R}T] \tag{63}$$

where $R$ is the concentration of Li in the reaction layer, and $K_-$ is a second characteristic rate. Here, $\dot{P}_A$ is the net rate of production of Li, in mols/second per unit reference area of the interface.

The net reaction rate vanishes when $\phi_A - \phi_F = \Delta\phi_0$, where $\Delta\phi_0$ is known as the 'rest potential'. The rest potential can be calculated from eq. (63), and is found to be related to the concentrations of electrons, ions and Li atoms in the reaction layer by

$$\Delta\phi_0 = \frac{\bar{R}T}{eN_a}\left(\log \frac{K_+}{K_-} + \log \frac{QR^+}{R}\right) \tag{64}$$

This relation is known as the Nernst equation [Hamann *et al.* (1998)]. With the aid of this result, equation (63) is usually re-written in a form that relates the electric current flow associated with the reaction to the 'overpotential' of the interface (i.e. the difference between the potential drop across the interface and the rest potential). We define the overpotential as $\Delta\phi_{AF} = \phi_A - \phi_F - \Delta\phi_0$, and note that the current flow into the interface per unit reference area is $I_A = -eN_A \dot{P}_A$. Then, eqs (63) and (64) can be combined to

$$I_A = I_0 \left\{ \exp\left(\frac{\alpha e N_a \Delta\phi_{AF}}{\bar{R}T}\right) - \exp\left(-\frac{(1-\alpha)eN_a \Delta\phi_{AF}}{\bar{R}T}\right) \right\} \tag{65}$$

where

$$I_0 = eN_a \left(K_+ Q R^+\right)^\alpha \left(K_- R\right)^{1-\alpha} \tag{66}$$

is the 'exchange current density' (the equal and opposite currents that flow across the interface when $\Delta\phi_{FA} = 0$). It should be noted that the sign convention for positive current flow is chosen so that current flows *into* the interface from the anode, i.e. electrons flow *out* of the reaction layer. With this convention, a positive flow of current corresponds to a net reaction in the direction $\text{Li} \to \text{Li}^+ + e^-$. Equations (65) and (66) are the standard form for the Butler-Volmer equation [Hamann *et al.* (1998)]. However, it should be noted that in our formulation we apply the Butler-Volmer equation across only the Stern layer, not (as is common in textbook treatments) across the entire double layer (including the diffuse regions). In this regard our model is identical to Biesheuvel *et al* (2009), who give a more detailed discussion of the issue.

We next investigate how stress in the electrode and electrolyte influences the interface reaction. It is important to note that surface concentrations of reactants $Q, R^+$ and reaction product $R$ in the interfacial layer are not equal to the concentrations in the bulk of the solid. The concentrations in the surface layer and in the bulk are related, however. Here we have assumed that the bulk of the solid is in equilibrium with the reaction layer, so the electrochemical potentials in the bulk and in the surface layer are related by eqs. (56). If we take the chemical potentials of Li, electrons, and Li$^+$ in the bulk and electrolyte to be

$$\frac{\partial \psi_0}{\partial c} = \bar{R}T \log \frac{\gamma c}{c_0} \qquad \frac{\partial \psi_0}{\partial q} = \bar{R}T \log \frac{\gamma_q q}{q_0} \qquad \frac{\partial \psi^F}{\partial \rho^+} = \bar{R}T \log \frac{\gamma^+ \rho^+}{\rho_0^+} \tag{67}$$

where $\gamma, \gamma_q, \gamma^+$ are activity coefficients and $c_0, q_0, \rho_0^+$ are reference concentrations for the bulk solid, then, equilibrium between the bulk and surface requires that

$$\begin{aligned}
\bar{R}T \log \frac{\Gamma^+ R_A^+}{R_0^+} &= \bar{R}T \log \frac{\gamma^+ \rho_A^+}{\rho_0^+} - N_k F_{ik} J \tilde{\sigma}_{ij}^F F_{lj}^{-1} N_l \Omega N_a \\
\bar{R}T \log \frac{\Gamma_Q Q}{Q_0} &= \bar{R}T \log \frac{\gamma_q q}{q_0} \\
\bar{R}T \log \frac{\Gamma R_A}{R_0} &= \bar{R}T \log \frac{\gamma c_A}{c_0} - \frac{\beta^2(c)}{2\rho_{0A}} \frac{\partial S_{ijkl}}{\partial c} \sigma_{ij} \sigma_{kl} - \frac{\sigma_{kk}}{3\rho_{0A}} \frac{\partial \beta}{\partial c}
\end{aligned} \tag{68}$$

These relations can be used to express the surface concentrations $Q, R^+ \, R$ in equation (63) by the bulk concentrations $\rho^+, q$ and $\rho$. Following the same steps that lead to (65) and (66), we obtain a modified form for the Butler-Volmer equation, which is expressed in terms of the bulk concentrations in the electrode and electrolyte just adjacent to the reaction layer

$$I_A = i_0 \left\{ \exp\left(\frac{\alpha e N_a (\phi_A - \phi_F - \Delta\phi_0)}{\bar{R}T}\right) - \exp\left(-\frac{(1-\alpha) e N_a (\phi_A - \phi_F - \Delta\phi_0)}{\bar{R}T}\right) \right\} \tag{69}$$

Here, the exchange current density $i_0$ is now a function of the bulk concentrations

$$i_0 = eN_a \left(k_+ q\rho^+\right)^\alpha \left(k_- \rho\right)^{1-\alpha} \tag{70}$$

with $k_+, k_-$ a pair of new reaction rate coefficients, which depend on $K_+, K_-$ and the activities and reference concentrations of the species in the bulk and the reaction layer. In addition, the rest potential $\Delta\phi_0$ in (69) is a function of the bulk concentrations and stress in the solid and in the electrolyte just adjacent to the reaction layer

$$\Delta\phi_0 = \frac{\bar{R}T}{eN_a}\left(\log\frac{k_+}{k_-} + \log\frac{q\rho^+}{c}\right) + \frac{1}{eN_a}\left(\frac{\beta^2(c)}{2\rho_{0A}}\frac{\partial S_{ijkl}}{\partial c}\sigma_{ij}\sigma_{kl} + \frac{\sigma_{kk}}{3\rho_{0A}}\frac{\partial \beta}{\partial c} - N_k F_{ik} J \tilde{\sigma}^F_{ij} F^{-1}_{lj} N_l \Omega N_a \right) \tag{71}$$

It is helpful to reiterate the definitions of the variables in the terms involving stress: $\rho_{0A}$ is the number of mols of Si atoms per unit volume in a stress free Si anode containing no Li; $\beta(c)$ quantifies the volume of a stress free anode containing a molar concentration $c$ of Li; $\sigma_{ij}$ is the stress in the anode; $S_{ijkl}$ is the elastic compliance tensor of the anode; $\Omega$ is the volume of an Li atom; $N_i$ is a vector normal to the undeformed surface of the anode; $F_{ij}$ is the deformation gradient in the electrolyte; while $\tilde{\sigma}^F_{ij} = \sigma^{MF}_{ij} - \pi\delta_{ij}$, where $\sigma^{MF}_{ij}$ and $\pi$ are the Maxwell stress and pressure in the electrolyte, respectively. Again, it should be noted that field quantities appearing in eq. (71) are those just outside the Stern layer – they are not the bulk quantities far from the reaction layer.

Equation (71) can be regarded as a generalized Nernst equation, which relates the rest potential to the concentrations of reactants in the electrode and electrolyte. In the absence of stress, the expression reduces to the classical form. Stress may increase or decrease the rest potential, depending on its sign. For the Si/electrolyte interface, the rest potential for a stress free surface $\Delta\phi_0 > 0$. Consequently, at fixed concentration, a compressive stress in the anode will reduce the rest potential; while a tensile stress will increase it. experimental evidence demonstrating this behavior has been reported by Sethuraman *et al* (2010c). Their data will be compared with the predictions of our model in the next section.

Before proceeding, however, we show that the relationship between the rest potential and stress can be obtained by means of a more direct argument, which does not require a detailed model of the interface reaction, or a specific form for the concentration dependence of free energy. If the system is in equilibrium, then the rate of change of free energy given in (28) and the following equations must *vanish* for all possible combinations of fluxes. The last integral in (32) then immediately relates the electrochemical potentials of the reactants and reaction products in the interfacial layer as

$$M_{A+} + M_{AQ} = M_{AR} \tag{72}$$

We can express this result in terms of the electrochemical potentials in the bulk anode and electrolyte using eqs. (56), and use the expression for the electrochemical potential of Li in the anode from eq. (45) to see immediately that the electrical potential difference across the interfacial layer at the anode is

$$\phi_A - \phi_F = \Delta\phi_{0chem} + \Delta\phi_{0mech}$$

$$\Delta\phi_{0chem} = \frac{1}{eN_a}\left\{\frac{\partial \psi^F}{\partial \rho^+} + \rho_{0A}\frac{\partial \psi_0}{\partial q} - \frac{\partial \psi_0}{\partial c}\right\} \quad (73)$$

$$\Delta\phi_{0mech} = \frac{1}{eN_a}\left\{\frac{\beta^2(c)}{2\rho_{0A}}\frac{\partial S_{ijkl}}{\partial c}\sigma_{ij}\sigma_{kl} + \frac{\sigma_{kk}}{3\rho_{0A}}\frac{\partial \beta}{\partial c} - N_k F_{ik} J \tilde{\sigma}_{ij}^F F_{lj}^{-1} N_l \Omega N_a\right\}$$

where $\Delta\phi_{0chem}, \Delta\phi_{0mech}$ represent chemical and mechanical contributions to the rest potential. This result is equivalent to (71).

Finally, we note that the flow of current across the cathode/electrolyte interface can be treated in a similar manner. The reaction within the Stern layer at the cathode can be described by (63), with $\phi_A - \phi_F$ replaced by $\phi_C - \phi_F$. The surface concentrations can then be expressed in terms of the bulk concentrations of free electrons and Li$^+$. Omitting the details, the current flow into the reaction layer can be expressed as

$$I_C = i_{0C}\left\{\exp\left(\frac{\alpha e N_a(\phi_C - \phi_F - \Delta\phi_{0C})}{\bar{R}T}\right) - \exp\left(-\frac{(1-\alpha)eN_a(\phi_C - \phi_F - \Delta\phi_{0C})}{\bar{R}T}\right)\right\} \quad (74)$$

where the rest potential for the cathode is

$$\Delta\phi_{0C} = \frac{\bar{R}T}{eN_a}\left\{\log\frac{k_{C+}}{k_{C-}} + \log\frac{q\rho^+}{R_0}\right\} + \frac{\Omega}{e}\left\{F_{ik}^C N_k \tilde{\sigma}_{ij}^C JF_{lj}^{C-1} N_l - \psi^C + \Psi^C \kappa - N_k F_{ik}^F \tilde{\sigma}_{ij}^F F_{lj}^{F-1} N_l\right\} \quad (75)$$

## 5. Model of a half-cell with a thin-film Si anode

Sethuraman *et al* (2010b) recently reported direct measurements of stress in a Li-ion half-cell with an anode made from a thin film of Si. Their experimental apparatus is shown in Fig. 2, and will be described in more detail in the next section. With a view to comparing the predictions of our model with their experimental data, we apply the model to a 1-D approximation of their specimen. Their cell will be idealized as a film of Si with initial thickness $h_{A0}$ on a rigid substrate; a fluid electrolyte layer with depth $h_F$ and a solid Li cathode (see Fig 1). The anode is an isotropic, elastic-plastic solid, with Young's modulus $E$, Poisson's ratio $\nu$, flow stress $\sigma_0$ and characteristic strain rate and stress exponent $\dot{d}_0, m$. For simplicity, we shall assume that both the anode and cathode are perfect conductors, and neglect the thermal energy of electrons in the bulk of the electrodes. In addition, we assume that the cathode is stress free. A potentiostat maintains a fixed flow of electric current $I^*$ per unit area from the anode to the cathode.

The deformation gradient in the Si anode can be decomposed into contributions from elastic deformation, volumetric expansion due to Li insertion, and plastic deformation as

$$\mathbf{F} = \begin{bmatrix} 1+\varepsilon & 0 & 0 \\ 0 & 1+\varepsilon & 0 \\ 0 & 0 & 1-2\nu\varepsilon/(1+\nu) \end{bmatrix} \begin{bmatrix} \beta^{1/3} & 0 & 0 \\ 0 & \beta^{1/3} & 0 \\ 0 & 0 & \beta^{1/3} \end{bmatrix} \begin{bmatrix} 1/\sqrt{\lambda_p} & 0 & 0 \\ 0 & 1/\sqrt{\lambda_p} & 0 \\ 0 & 0 & \lambda_p \end{bmatrix} \quad (76)$$

where $\lambda_p$ represents out-of-plane plastic stretching of the film; $\beta(c)$ represents the volumetric expansion due to Li insertion, and $\varepsilon \ll 1$ is the elastic in-plane stretch. The substrate prevents the thin film anode from stretching in the horizontal plane, so that

$$F_{11} = F_{22} = (1+\varepsilon)\beta^{1/3}/\sqrt{\lambda_p} = 1 \quad (77)$$

Taking a time derivative yields

$$\frac{\dot{\beta}}{3\beta} + \dot{\varepsilon} - \frac{\dot{\lambda}_p}{2\lambda_p} = 0 \quad (78)$$

The plastic part of the velocity gradient in the anode can be approximated as

$$L_{ij}^p = F_{pn}^e F_{ni}^c \dot{F}_{ij}^p F_{jl}^{p-1} F_{lk}^{c-1} F_{kq}^{e-1} \approx \begin{bmatrix} -\dot{\lambda}_p/2\lambda_p & 0 & 0 \\ 0 & -\dot{\lambda}_p/2\lambda_p & 0 \\ 0 & 0 & \dot{\lambda}_p/\lambda_p \end{bmatrix} \quad (79)$$

The Cauchy stress in the anode has nonzero components $\sigma_{11} = \sigma_{22} = \sigma$. The Kirchoff stress follows as $\tau_{11} = \tau_{22} \approx \beta\sigma$, with deviatoric and effective stresses $\tau_{11}^D = \tau_{22}^D = \beta\sigma/3$ $\tau_{33}^D = -2\beta\sigma/3$, $\tau_e = |\beta\sigma|$. The elastic and plastic constitutive equations (39) and (43) then yield

$$\beta\sigma = \frac{E}{1-\nu}\varepsilon \qquad -\frac{\dot{\lambda}_p}{\lambda_p} = \dot{d}_0 \left( \frac{\beta|\sigma|}{\sigma_0} - 1 \right)^m \frac{\sigma}{|\sigma|} \quad (80)$$

Diffusion of Li through the anode is governed by eqs. (2), (46) and the second of (49). These reduce to

$$j_3 = -\rho_{0A} c D \frac{\partial \mu_A}{\partial X_3} \qquad \rho_{0A} \frac{\partial c}{\partial t} = -\frac{\partial j_3}{\partial X_3} \qquad \mu_A = \frac{\partial \psi_0}{\partial c} - \frac{\beta^2 \sigma^2}{\rho_{0A}} \frac{\partial}{\partial c}\left(\frac{1-\nu}{E}\right) - \frac{2\beta\sigma}{3\rho_{0A}} \frac{1}{\beta} \frac{\partial \beta}{\partial c} \quad (81)$$

We shall take the stress-free chemical potential of Li in the anode to be

$$\frac{\partial \psi_0}{\partial c} = \bar{R}T \log \frac{\gamma c}{c_0} \quad (82)$$

where $\bar{R}$ is the gas constant, $T$ is the temperature, $\gamma$ is the activity coefficient of Li in Si and $c_0$ is a reference concentration.

Since we assume that the anode is a perfect conductor, the electric potential $\phi_A$ is uniform in the anode and the electric field vanishes. The electron density is at equilibrium everywhere in the interior of the anode, and a surface layer of charge with density $Q_A$ develops at the anode/electrolyte interface.

Diffusion of Li$^+$ through the electrolyte is governed by equations (6), (51) and (52). These reduce to

$$j_3^+ = -\rho^+ D_+ \frac{\partial \mu_+}{\partial X_3} \qquad \frac{\partial \rho^+}{\partial t} = -\frac{\partial j_3^+}{\partial X_3} \qquad \mu^+ = \frac{\partial \psi^F}{\partial \rho^+} + eN_a\phi \tag{83}$$

where we take the chemical potential of Li$^+$ in the electrolyte to be

$$\frac{\partial \psi^F}{\partial \rho^+} = \bar{R}T \log \frac{\gamma^+ \rho^+}{\rho_0^+} \tag{84}$$

The electric field and electric potential in the electrolyte are related to the Li$^+$ concentration by the second of eq. (17), which reduces to

$$\frac{\partial H_3}{\partial X_3} - \frac{eN_a\left(\rho^+ - \rho_0^+\right)}{\varepsilon_0} = 0 \qquad \frac{\partial \phi}{\partial X_3} = -H_3 \tag{85}$$

The electrolyte is incompressible and boundary conditions prevent the electrolyte from stretching laterally. Consequently, the electrolyte remains rigid, with $F_{ij} = \delta_{ij}$. The pressure $\pi$ in the electrolyte must satisfy the equilibrium equation (21), which reduces to

$$\frac{\partial \pi}{\partial X_3} = eN_a(\rho^+ - \rho_0^+)H_3 \tag{86}$$

The cathode is free of stress, and has a uniform electric potential $\phi_C$ and vanishing electric field. A surface layer of charge with density $Q_C$ develops at the cathode/electrolyte interface.

Boundary conditions at the interface between the anode and electrolyte are listed in eq. (18), as well as sections 4.5 and 4.6. We take the chemical potential of Li and Li$^+$ in the reaction layers to have the form given in eq. (60). In addition, the electrolyte within the Stern layer is prevented from deforming, so that $F_{ij}^* = \delta_{ij}$. Equations (18) then relate the rate of change of potential difference across the interface to the concentration of electrons $Q$ and Li$^+$, $R_A^+$, in the interface as

$$\frac{d}{dt}(\phi_A - \phi_{FA}) = -\frac{eN_a a}{\varepsilon_0}\frac{dQ_A}{dt} \qquad \frac{dH_3^{FA}}{dt} = \frac{eN_a a}{\varepsilon_0}\left(\frac{dR_A^+}{dt} - \frac{dQ_A}{dt}\right) \tag{87}$$

where $\phi_{FA}, H_3^{FA}$ denote the electric potential and electric field in the electrolyte just outside the Stern layer, and $a$ is the thickness of the layer. The third conservation law in eq (8) gives the rate of change of electron density in the Stern layer

$$\frac{dQ_A}{dt} = \frac{I^* + I_A}{eN_a} \tag{88}$$

where $I_A$ is the electric current at the interface, given in eq. (65).

The reactants and reaction products within the reaction layer must be in equilibrium with the adjacent bulk electrode and electrolyte. Consequently, the concentrations at the anode/electrolyte interface are related by eqs. (68), which, for a biaxial stress state reduce to

$$\log \frac{\gamma^+ \rho_A^+}{\rho_0^+} = \log \frac{\Gamma^+ R_A^+}{R_0^+} \qquad \log \frac{\Gamma R_A}{R_0} = \log \frac{\gamma c_A}{c_0} - \frac{\beta^2 \sigma^2}{\overline{RT}\rho_{0A}} \frac{\partial}{\partial c}\left(\frac{1-\nu}{E}\right) - \frac{2\beta\sigma}{3\overline{RT}\rho_{0A}} \frac{1}{\beta}\frac{\partial \beta}{\partial c} \tag{89}$$

The term in eq. (68) involving the stress in the electrolyte vanishes in this problem, because charge is assumed to be concentrated in the Stern layer. Finally, the first two conservation laws in eq. (8) give boundary conditions for the flux of Li and Li$^+$ at the interface

$$j_3 = \frac{\partial R_A}{\partial t} + \frac{I_A}{eN_a} \qquad j_3^+ = \frac{\partial R_A^+}{\partial t} - \frac{I_A}{eN_a} \tag{90}$$

Similar boundary conditions hold at the cathode, but these will not be listed in detail.

Finally, equations (78)-(82), eqs. (83)-(85) may be combined to yield a pair of equations governing the evolution of Li concentration and stress in the anode

$$\frac{\partial c}{\partial t} = -D \frac{\partial j_3}{\partial X_3} = D \frac{\partial}{\partial X_3} \left\{ \overline{RT} \frac{\partial c}{\partial X_3} - c \frac{\partial}{\partial X_3}\left[\frac{\beta^2 \sigma^2}{\rho_{0A}} \frac{\partial}{\partial c}\left(\frac{1-\nu}{E}\right) - \frac{2\beta\sigma}{3\rho_{0A}} \frac{1}{\beta}\frac{\partial \beta}{\partial c}\right]\right\}$$

$$\frac{\partial}{\partial t}\left(\frac{1-\nu}{E}\beta\sigma\right) = -\frac{1}{3\beta}\frac{\partial \beta}{\partial c}\frac{\partial c}{\partial t} - \dot{d}_0 \left(\frac{\beta|\sigma|}{\sigma_0} - 1\right)^m \frac{\sigma}{|\sigma|} \tag{91}$$

while eqs. (83)-(85) can be combined to relate the Li$^+$ concentration and electric potential in the electrolyte as follows

$$\frac{\partial \rho^+}{\partial t} = D\frac{\partial}{\partial X_3}\left\{\overline{RT}\frac{\partial \rho^+}{\partial X_3} + \rho^+ eN_a \frac{\partial \phi}{\partial X_3}\right\} \qquad \frac{\partial^2}{\partial X_3^2}\frac{\partial \phi}{\partial t} = -eN_a \frac{\partial \rho^+}{\partial t} \tag{92}$$

The boundary conditions at the anode can be combined to

$$\frac{d}{dt}(\phi_A - \phi_{FA}) = -(I_A + I^*)a/\varepsilon_0 \qquad \frac{dH_3^{FA}}{dt} = (2I_A - I^* - eN_a j_3^+)/\varepsilon_0$$

$$j_3 = \frac{\partial R_A}{\partial t} + \frac{I_A}{eN_a} \qquad j_3^+ = -\frac{\gamma^+ R_0^+}{\Gamma^+ \rho_0^+}\frac{\partial \rho_A^+}{\partial t} + \frac{I_A}{eN_a}$$

(93)

where $\partial R_A/\partial t$ can be expressed in terms of $\partial c_A/\partial t$ and the time derivative of the stress using eq. (89). Similarly, at the cathode,

$$\frac{d}{dt}(\phi_C - \phi_{FC}) = -(I_C - I^*)a/\varepsilon_0 \qquad \frac{dH_3^{FC}}{dt} = (2I_C + I^* + eN_a j_3^+)/\varepsilon_0$$

$$j_3^+ = \frac{\gamma^+ R_0^+}{\Gamma^+ \rho_0^+}\frac{\partial \rho_C^+}{\partial t} + \frac{I_C}{eN_a}$$

(94)

Equations (91)-(94), together with the Butler-Volmer equation in (69) form a complete set of equations and boundary conditions governing the stress, electric potential, and concentration fields in the half-cell. The equations can only be solved numerically, which will not be attempted here.

Instead, we proceed with an approximate solution, by assuming that (i) diffusion of Li in the anode and Li$^+$ through the electrolyte are both much faster than the rates of interface reaction in the Stern layers; and (ii) the activities and reference concentrations in the electrolyte and Stern layer satisfy $\gamma^+ R_0^+ \gg \Gamma^+ \rho_0^+$. Under these conditions, the stress and concentration fields in the anode are uniform, the density of free electrons, $Q$, and Li$^+$, $R^+$, in the reaction layers are equal, and the variations of Li$^+$ concentration and electric field in the bulk of the electrolyte may be neglected. The spatial derivatives in (91) and (92) may then be eliminated, yielding a system of ODEs governing the rate of change of electric potential $\phi_A, \phi_F$ in the anode and electrolyte, the stress $\sigma$ and Li concentration $c$ in the anode, and the concentrations $R_A^+, R_C^+$ of Li$^+$ in the Stern layers at the Anode and Cathode, respectively

$$\frac{d}{dt}(\phi_A - \phi_F) = -(I_A + I^*)a/\varepsilon_0$$

$$\frac{d}{dt}R_A^+ = -(I_A + I^*)a/(eN_a\varepsilon_0)$$

$$\frac{\partial}{\partial t}\left(\frac{1-\nu}{E}\beta\sigma\right) = -\frac{1}{3\beta}\frac{\partial \beta}{\partial c}\frac{dc}{dt} - \dot{d}_0\left(\frac{\beta|\sigma|}{\sigma_0} - 1\right)^m \frac{\sigma}{|\sigma|}$$

$$\frac{dc}{dt} = -\frac{I_A}{\rho_{0A}eN_a h_{A0}}$$

$$\frac{d}{dt}(\phi_C - \phi_F) = -(I_C - I^*)a/\varepsilon_0$$

$$\frac{d}{dt}R_C^+ = -(I_C - I^*)a/(eN_a\varepsilon_0)$$

(95)

We take the current flow at the anode to be given by

$$I_A = k_A R_A^{+2\alpha} c^{1-\alpha} \left\{ \exp\left( \frac{\alpha e N_a (\phi_A - \phi_F - \Delta\phi_{0A})}{\overline{R}T} \right) - \exp\left( \frac{(1-\alpha) e N_a (\phi_A - \phi_F - \Delta\phi_{0A})}{\overline{R}T} \right) \right\} \quad (96)$$

$$\Delta\phi_{0A} = \frac{\overline{R}T}{eN_a} \left( \Theta(c) + \log\frac{\left(R_A^+\right)^2}{c} \right) + \frac{1}{eN_a} \left\{ \frac{\beta^2 \sigma^2}{\rho_{0A}} \frac{\partial}{\partial c}\left(\frac{1-\nu}{E}\right) + \frac{2\beta\sigma}{3\rho_{0A}} \frac{1}{\beta} \frac{\partial\beta}{\partial c} \right\} \quad (97)$$

where $\Theta(c) = \log K_A^+ / k_{A-}$ must be fit to experimental data. A similar expression applies to the current flow at the cathode.

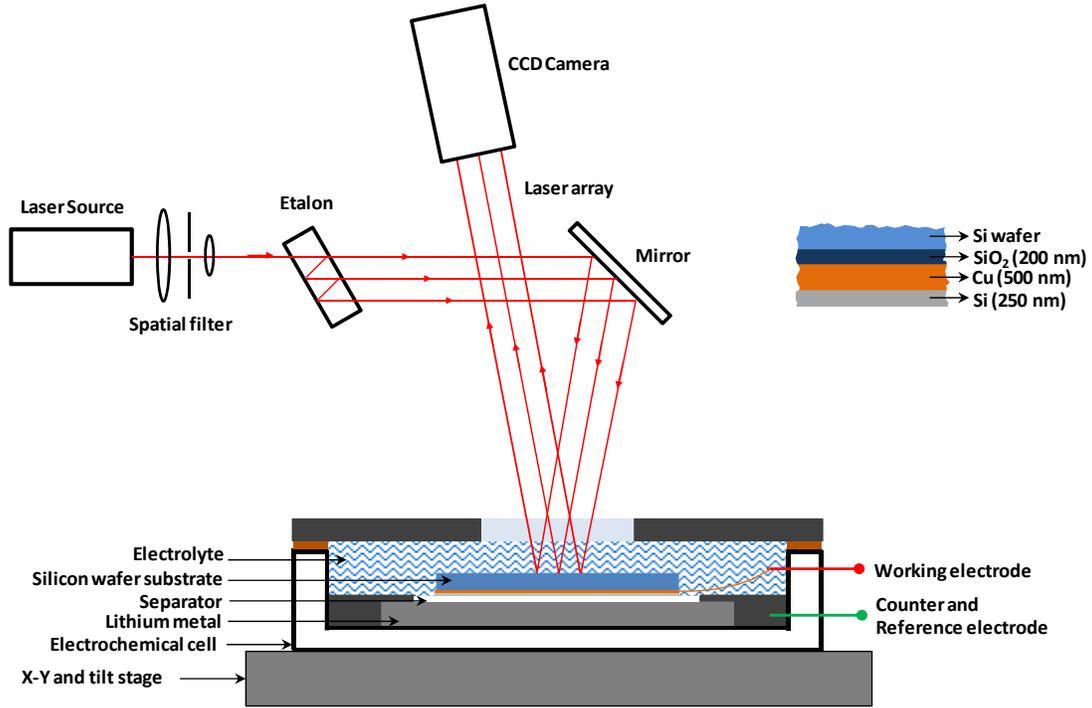

*Figure 2: Schematic illustration of the home-made electrochemical cell is shown along with the multi-beam-optical sensor setup to measure substrate curvature. Note that the schematic is not drawn to scale. In this two-electrode configuration, Si thin film is the working electrode, and Li metal is the counter and reference electrode. Above right: the layered configuration of the working electrode on Si-wafer substrate is shown. Figure is from Sethuraman et al. (2010c).*

The predictions of (95) will be compared with experimental data in the next section. In these calculations, we take $\partial\beta / \partial c$ to be constant, and assume that the Young's modulus $E$ and flow stress of the anode vary linearly with Li concentration $c$. Similarly, $\Theta(c)$ is taken to be linear. The rest potential at the cathode is assumed to be zero, and we assume that the exchange current density at the cathode greatly exceeds that at the anode. Under these conditions the electric potential in the electrolyte is equal to that of the cathode.

**6. Comparison to experiment**

We conclude by comparing the predictions of the model described in the preceding section to the experimental data obtained by Sethuraman *et al*. (2010c). Only a brief summary of the experiment will be given here, since a detailed description is available in the above reference.

The measurements were made on the model electrochemical half-cell illustrated in Fig 2. The working electrode consists of a 250nm thick film of amorphous Si. The Si is deposited (by sputtering) on a 500nm thick layer of Cu, which serves as a current collector. The reference electrode is a 50.8mm and 1.5mm thick disk of solid Li. The electrolyte consists of 1.2 M lithium hexafluoro-phosphate in 1:2 (vol. %) ethylene carbonate:diethyl carbonate with 10% fluoroethylene carbonate additive. To detect the stress in the working electrode, the Cu layer and overlying Si film are deposited on a thin film substrate, consisting of single crystal Si wafer with (111) orientation, 50.8mm diameter and 450 micron thickness. The curvature of the substrate is measured using a multi-beam optical sensor (MOS) during operation of the cell. The biaxial Cauchy stress in the film is then calculated from the curvature using the Stoney equation

$$\sigma = \frac{E_s h_s^2 \kappa_s}{6 h_f (1-\upsilon_s)} \qquad (98)$$

where $E_s, \nu_s, h_s$ are the Young's modulus, Poisson's ratio and thickness of the substrate; $h_f$ is the thickness of the amorphous Si electrode, and $\kappa_s$ is the substrate curvature. The thickness of the electrode varies as the material expands during charging: its thickness was estimated as

$$h_f = h_{A0}(1 + 2.7 z / z_{\max}) \qquad (99)$$

where $h_{A0}$ is the initial thickness of the amorphous Si film; $z$ is the charge per unit initial mass in the electrode, and $z_{\max} = 3579$ mAh/g is the maximum charge capacity. It should be noted that *ca.* 0.25 GPa growth stress is present in the film prior to lithiation, so that (98) should be interpreted as predicting the *change* in stress during lithiation, not as the absolute value of the stress in the film. The initial stress in the film was determined by measuring the change in curvature of the substrate during sputtering.

The variation of stress and electric potential were monitored during the experiment. Two sets of tests were conducted. In the first, the working electrode was first lithiated at a constant current density of 0.125 Am$^{-2}$ until the potential decreased to 0.01V. The electrode was then delithiated at the same current. The resulting cycles of stress and electric potential are plotted as functions of the charge in Fig. 3.

These data were used to calibrate the constitutive equations given in the preceding section. Parameter values and initial conditions were either determined from literature data, or selected to yield the best fit to the data as listed in Table 1. The predicted cycles of stress and electric potential are shown in Fig. 3. The model clearly reproduces both qualitatively and quantitatively the main features of the experimental results. During initial lithiation, compressive stresses are induced in the film. These stresses exceed the yield stress of the film, causing it to permanently increase in thickness, while simultaneously reducing its in-plane dimensions to maintain its volume.

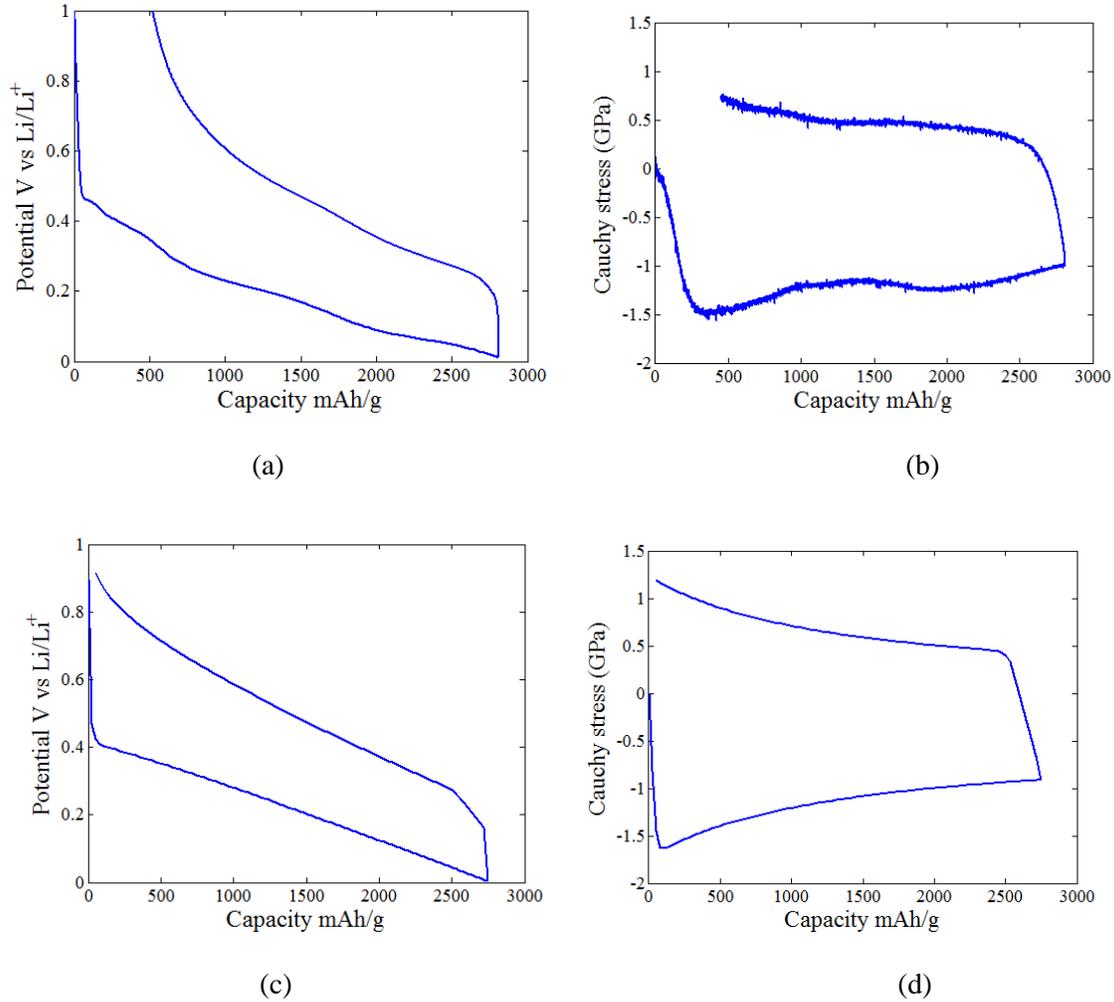

*Figure 3: Comparison of measure (a,b) and predicted (c,d) cycles of potential and stress during initial lithiation and delithiation of amorphous silicon thin-film electrode. Figs (a,c) show the potential history, while (c,d) show the stress. Experimental data is replotted from Sethuraman et al. (2010c).*

When the film is subsequently de-lithiated, the film initially unloads elastically, and quickly develops a tensile stress, as a consequence of the plastic flow during lithiation. Further de-lithiation eventually causes the film to yield in tension. The effects of lithiation and de-lithiation are identical to the effects of thermal cycling. There are only three notable discrepancies between theory and experiment: First, the model underestimates the transient drop in potential at the start of lithiation; second, the model predicts a more rapid increase in stress during lithiation than is observed in the experiment; and thirdly, the model underestimates the rise in potential towards the end of the delithiation cycle. These discrepancies can all be attributed to electrochemical processes that are not considered in our model. In particular, we have neglected the solid electrolyte interphase (SEI) layer that forms at the surface of the Si electrode during the first charging cycle. In addition, our model neglects capacity fade caused by parasitic side reactions.

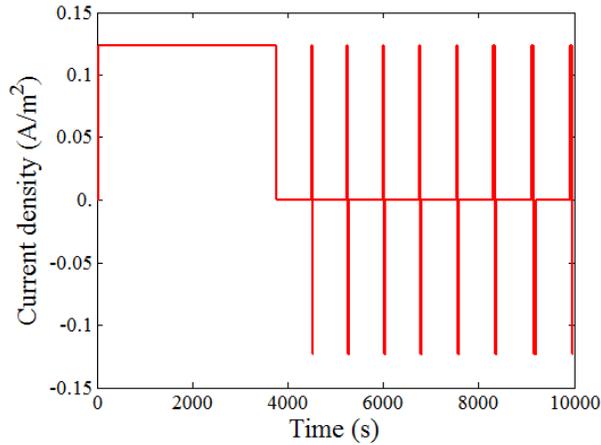

*Fig. 4: History of current density during cyclic lithiation and de-lithiation of Si thin film electrode. A positive current density corresponds to lithiation.*

Sethuraman *et al.* (2010c) also conducted a second series of experiments that were designed to detect the stress-potential coupling discussed in Section 4.6 of this paper. In these experiments, the electrode was first lithiated to a desired potential, $V$ ($1.2 \leqslant V \leqslant 0.010$ V *vs.* Li/Li$^+$), then delithiated for a small period of time, $t_d$, followed by an open-circuit-potential relaxation for one hour. After the open circuit relaxation step, the electrode was re-lithiated to the same initial potential $V$, and delithiated for an incrementally larger duration than before, *i.e.*, for a period $t_d + \Delta t$ followed by an hour-long open-circuit relaxation. A typical history of current density is shown in Fig. 4. The increase in delithiation time allows for incrementally larger elastic unloading thereby taking the electrode stress to a new value. This sequence of steps was repeated with progressively longer delithiation time, allowing the sample to relax towards a steady state potential at several values of stress between the compressive and tensile flow stress, resulting in a total stress change of over 1 GPa.

Fig. 5 compares the measured and predicted cycles of stress and potential induced by the history of current shown in Fig 4. The model again clearly captures the main features of the experiment both qualitatively and quantitatively. Two discrepancies between theory and experiment are notable: first, the variation of potential and stress during the initial 3500s charging cycle do not match. The measured history of stress and potential during the initial transient are not repeatable, however. This variation is likely to be caused by the formation of the SEI layer, which is not considered in our calculations. Secondly, the experiments show a systematic increase in stress throughout each 1-hour long zero current hold. The stress continues to increase during the zero current phase even after it reaches 0.5GPa tensile stress, so this behavior is unlikely to be caused by stress relaxation. Instead, it is likely that side reactions cause a loss of capacity, with an accompanying increase in tensile stress. More experiments are required to fully explain this behavior, however.

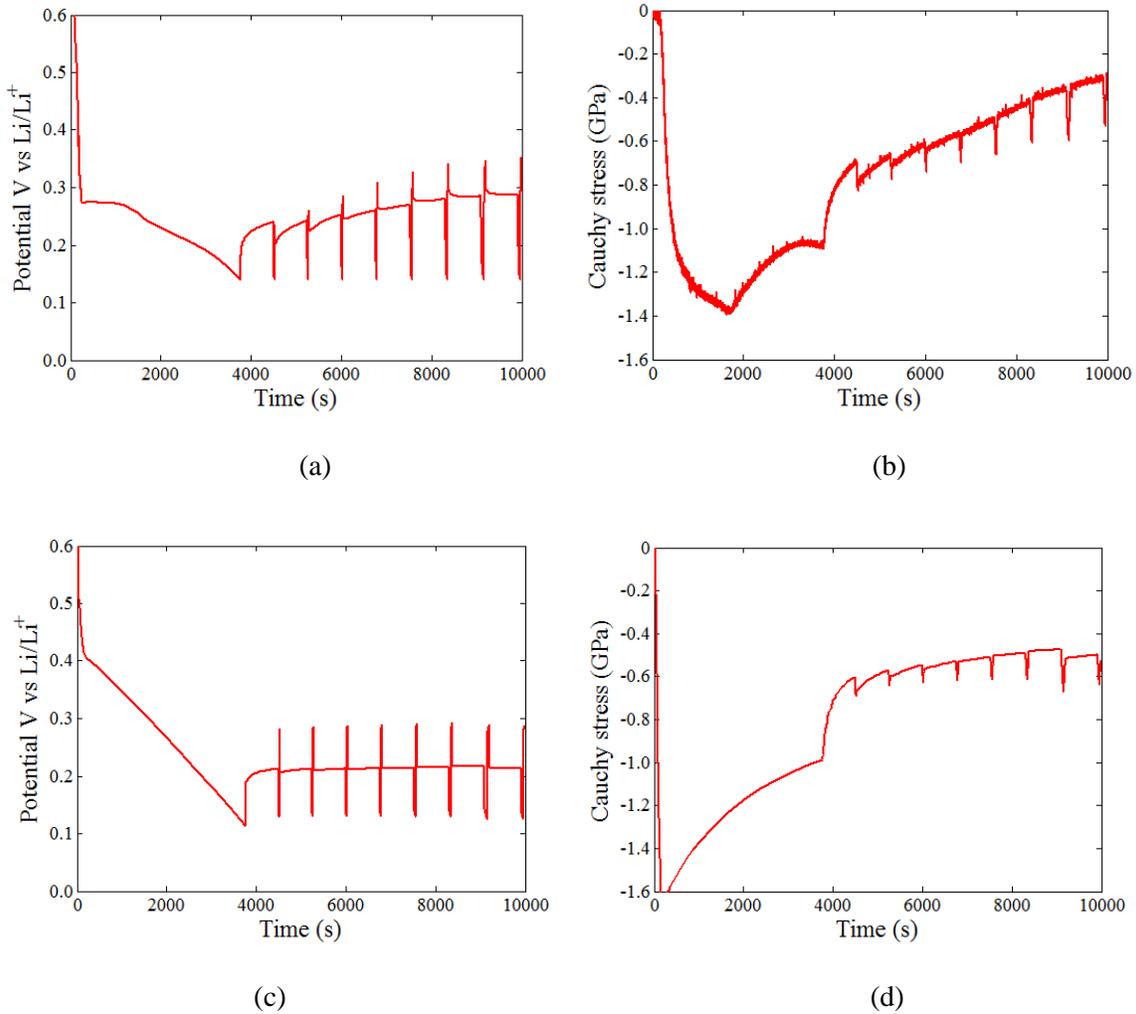

*Fig. 5: Measured (a,b) and predicted (c,d) variation of potential (a,c) and stress (b,d) in an Si electrode during the cycles of current shown in Fig 4.*

## 7. Conclusions

We have described a continuum model of deformation, diffusion, stress, electrostatic fields, and electrochemical reactions in a Li-ion half-cell. Our model extends earlier work by accounting rigorously for the effects of finite deformations, for inelastic deformation, and the coupling between mechanical stress and electrochemical reactions at the interface. We have listed the complete set of field equations, boundary conditions, and constitutive relations that govern the various processes that occur in the cell, but three results are of particular interest.

Firstly, the elastic-plastic stress-strain relations for the anode are given in eqs (39) and (43). These are identical to the usual finite deformation stress-strain relations for an elastic-viscoplastic solid, except for the factor $\beta$, which accounts for the change in volume of the solid due to Li insertion. Secondly, diffusion of Li through the anode is governed by the diffusion equation in eq. (49), where the chemical potential of Li is given in eq. (46). Again, except for the factor $\beta$ multiplying the second term in the chemical potential, the finite strain result is identical to the small-strain Larché-Cahn chemical potential (Larché and Cahn, 1978), provided that Cauchy stress is used as the stress measure. Finally, thermodynamic considerations indicate that stress has a direct influence on the rest potential for the electrode-electrolyte interfaces. The effects of stress can be taken into account by modifying the Nernst equation for the rest potential of an interface as outlined in eq. (71).

Our finite-strain elastic-plastic model was used to predict the evolution of stress and potential for a one-dimension Li-ion half-cell with a Si anode. A rate dependent viscoplastic constitutive equation with parameters listed in Table 1 was used to model stress relaxation due to plastic flow in the Si. The predicted cycles of stress and potential were shown to be in good agreement with experimental measurements.

## 8. Acknowledgements

This work was supported in part by the Materials Research, Science and Engineering Center (MRSEC) sponsored by the United States National Science Foundation under contract no. DMR0520651; by the Rhode Island Science and Technology Advisory Council under grant no. RIRA 2010-26 (PG); and by the General Motors Collaborative Research Laboratory on Computational Materials Research (AFB).

**APPENDIX A**

Here, we derive the expression for the rate of change of free energy in the half-cell given in eqs. (28)-(33). We begin by taking the time derivative of eq. (27), and adding constraints

$$\frac{\partial G}{\partial t} = \int_{V_{0A}} \left( \frac{\partial \psi^A}{\partial F_{ij}^e} \dot{F}_{ij}^e + \frac{\partial \psi^A}{\partial F_{ij}^p} \dot{F}_{ij}^p + \frac{\partial \psi^A}{\partial \rho} \dot{\rho} + \frac{\partial \psi^A}{\partial q} \dot{q} + \frac{\partial \psi^A}{\partial H_i} \dot{H}_i \right) dV$$

$$+ \int_{V_{0F}} \left( \frac{\partial \psi^F}{\partial F_{ij}} \dot{F}_{ij} + \frac{\partial \psi^F}{\partial \rho^+} \dot{\rho}^+ + \frac{\partial \psi^F}{\partial H_i} \dot{H}_i \right) dV$$

$$+ \int_{V_{0C}} \left( \frac{\partial \psi^C}{\partial F_{ij}} \dot{F}_{ij} + \frac{\partial \psi^C}{\partial q} \dot{q} + \frac{\partial \psi^C}{\partial H_i} \dot{H}_i \right) dV - \int_{A_{0Ce}} (\psi^C - \Psi^C \kappa) \Omega N_a j_C dA$$

$$+ \int_{A_{0Ae}} \left( \frac{\partial \Psi}{\partial F_{ij}^*} \dot{F}_{ij}^* + \frac{\partial \Psi}{\partial Q} \dot{Q} + \frac{\partial \Psi}{\partial R^+} \dot{R}^+ + \frac{\partial \Psi}{\partial H_i^*} \dot{H}_i^* \right) adA$$

$$+ \int_{A_{0Ce}} \left( \frac{\partial \Psi}{\partial F_{ij}^*} \dot{F}_{ij}^* + \frac{\partial \Psi}{\partial Q} \dot{Q} + \frac{\partial \Psi}{\partial R^+} \dot{R}^+ + \frac{\partial \Psi}{\partial H_i^*} \dot{H}_i^* \right) adA$$

$$- \int_{A_{Aext}} t_i \dot{x}_i dA - \int_{A_{Cext}} t_i \dot{x}_i dA - \Lambda \left( \int_{A_{0Aext}} i_{Aext} dA + \int_{A_{0Cext}} i_{Cext} dA \right)$$

$$- \int_{V_{A0}} \mu_{qA} \left( \dot{q} + \frac{\partial i_j}{\partial X_j} \right) dV - \int_{V_{A0}} \mu_A \left( \dot{\rho} + \frac{\partial j_j}{\partial X_j} \right) dV$$

$$- \int_{V_{0F}} \mu^+ \left( \dot{\rho}^+ + \frac{\partial j_j}{\partial X_j} \right) dV - \int_{V_{0C}} \mu_{qC} \left( \dot{q} + \frac{\partial i_j}{\partial X_j} \right) dV$$

$$+ \int_{V_{0A}} \left( \varepsilon_0 \dot{H}_j \frac{\partial \phi}{\partial X_j} - e N_a \dot{q} \phi \right) dV_0 + \int_{V_{0C}} \left( \varepsilon_0 \dot{H}_j \frac{\partial \phi}{\partial X_j} - e N_a \dot{q} \phi \right) dV_0$$

$$+ \int_{V_{F0}} \left( \varepsilon_0 \dot{H}_j \frac{\partial \phi}{\partial X_j} + e N_a \dot{\rho}^+ \phi \right) dV - \int_{A_{0Ae}} \varepsilon_0 \left( \dot{H}_j^A N_j \phi_A - \dot{H}_j^F N_j \phi_F \right) dA$$

$$- \int_{A_{0Ce}} \varepsilon_0 \left( \dot{H}_j^C N_j \phi_C - \dot{H}_j^F N_j \phi_F \right) dA$$

$$- \int_{V_F} \pi \frac{\partial \dot{x}_i}{\partial x_i} dV - \int_{A_{0Ae}} a \pi_{ij}^* J^* \dot{F}_{ik}^* F_{kj}^{*-1} dA - \int_{A_{0Ce}} a \pi_{ij}^* J^* \dot{F}_{ik}^* F_{kj}^{*-1} dA$$

$$- \int_{A_{0Ae}} M_{A+}(\dot{R}^+ + j^+ + \dot{P}_A) + M_{AR}(\dot{R} - j_A - \dot{P}_A) + M_{AQ}(\dot{Q} - i_{Ae} - \dot{P}_A) dA$$

$$- \int_{A_{0Ce}} M_{C+}(\dot{R}^+ + j^+ + \dot{P}_C) + M_{CR}(\dot{R} - j_C - \dot{P}_C) + M_{CQ}(\dot{Q} - i_{Ce} - \dot{P}_C) dA$$

(100)

where $\kappa$ is the mean curvature of the cathode/electrolyte surface, with sign convention chosen so that a concave surface has positive curvature. The first line of eq. (100) accounts for the rate of change of free energy of the anode; the second is that of the electrolyte. The third line is the rate of change of energy of the cathode, and the fourth and fifth lines are the rate of change of free energy of the reaction layer. The sixth line is the rate of work done by external tractions, and the constraint enforcing conservation of electrons flowing from one electrode to the other. The next two lines enforce conservation of electrons, Li and Li$^+$ mass, and volume of electrolyte. The following three lines are the variational form of Gauss law, while the last line enforces conservation of electrolyte volume, and conservation of electrons flowing out of the anode and cathode. Now note that, by integrating by parts,

$$
\begin{aligned}
&-\int_{V_{0A}} \mu_{qA}\left(\frac{\partial i_j}{\partial X_j}\right)dV - \int_{V_{0A}} \mu_A\left(\frac{\partial j_j}{\partial X_j}\right)dV - \int_{V_{0F}} \mu^+\left(\frac{\partial j_j}{\partial X_j}\right)dV - \int_{V_{0C}} \mu_{qC}\left(\frac{\partial i_j}{\partial X_j}\right)dV \\
&= -\int_{A_{0Aext}} \mu_{qA} i_{Aext} dA - \int_{A_{0Ae}} \mu_{qA} i_{Ae} dA - \int_{A_{0Ae}} \mu_A j_{Ae} dA + \int_{A_{0Ae}} \mu^+ j^+_{Ae} dA + \int_{A_{0Ce}} \mu^+ j^+_{Ce} dA \\
&\quad - \int_{A_{0Ce}} \mu_{qC} i_{Ce} dA - \int_{A_{0Cext}} \mu_{qC} i_{Cext} dA + \int_{V_{0A}} \frac{\partial \mu_{qA}}{\partial X_j} i_j dV + \int_{V_{0A}} \frac{\partial \mu_A}{\partial X_j} j_j dV + \int_{V_{0F}} \frac{\partial \mu^+}{\partial X_j} j^+_j dV + \int_{V_{0C}} \frac{\partial \mu_{qC}}{\partial X_j} i_j dV
\end{aligned}
\quad (101)
$$

In addition, note that eq. (18) implies that

$$\varepsilon_0 \dot{H}_i^F N_i = \varepsilon_0 \dot{H}_i^* N_i + e N_a \dot{R}^+ \qquad \varepsilon_0 H_i^A N_i = \varepsilon_0 H_i^* N_i + e N_a \dot{Q} \qquad (102)$$

at the anode/electrolyte interface, with a similar expression for the cathode. Substituting these relations, and substituting for the integrals involving external tractions using equation (26), eq. (100) can be rearranged in the form

$$\frac{\partial G}{\partial t} = \frac{\partial G^A}{\partial t} + \frac{\partial G^C}{\partial t} + \frac{\partial G^F}{\partial t}$$

$$+ \int_{A_{0Ae}} \left( \frac{\partial \Psi^A}{\partial F_{ij}^*} \dot{F}_{ij}^* + \frac{\partial \Psi^A}{\partial Q} \dot{Q} + \frac{\partial \Psi^A}{\partial R^+} \dot{R}^+ + \frac{\partial \Psi^A}{\partial H_i^*} \dot{H}_i^* \right) a \, dA$$

$$+ \int_{A_{0Ce}} \left( \frac{\partial \Psi^C}{\partial F_{ij}^*} \dot{F}_{ij}^* + \frac{\partial \Psi^C}{\partial Q} \dot{Q} + \frac{\partial \Psi^C}{\partial R^+} \dot{R}^+ + \frac{\partial \Psi^C}{\partial H_i^*} \dot{H}_i^* \right) dA$$

$$+ \int_{A_{0Ae}} (\tilde{\sigma}_{ij}^A \dot{x}_i^A - \tilde{\sigma}_{ij}^F \dot{x}_i^F) n_j \, dA + \int_{A_{0Ce}} (\tilde{\sigma}_{ij}^C \dot{x}_i^C - \tilde{\sigma}_{ij}^F \dot{x}_i^F) n_j \, dA$$

$$- \int_{A_{0Aext}} \mu_{qA} i_{Aext} \, dA - \int_{A_{0Ae}} \mu_{qA} i_{Ae} \, dA - \int_{A_{0Ae}} \mu_A j_{Ae} \, dA$$

$$+ \int_{A_{0Ae}} \mu^+ j_{Ae}^+ \, dA + \int_{A_{0Ce}} \mu^+ j_{Ce}^+ \, dA - \int_{A_{0Ce}} \mu_{qC} i_{Ce} \, dA - \int_{A_{0Cext}} \mu_{qC} i_{Cext} \, dA$$

$$- \int_{A_{0Ce}} (\psi^C - \Psi^C \kappa) \Omega_C N_a j_C \, dA - \Lambda \left( \int_{A_{0Aext}} i_{Aext} \, dA + \int_{A_{0Cext}} i_{Cext} \, dA \right)$$

$$- \int_{V_F} \pi \frac{\partial \dot{x}_i}{\partial x_i} dV - \int_{A_{0Ae}} a \pi_{ij}^* J^* \dot{F}_{ik}^* F_{kj}^{*-1} dA - \int_{A_{0Ce}} a \pi_{ij}^* J^* \dot{F}_{ik}^* F_{kj}^{*-1} dA$$

$$- \int_{A_{0Ae}} (\varepsilon_0 H_i^* N_i + e N_a \dot{Q}) \phi_A - (\varepsilon_0 \dot{H}_i^* N_i + e N_a \dot{R}^+) \phi_F \, dA$$

$$- \int_{A_{0Ce}} (\varepsilon_0 H_i^* N_i + e N_a \dot{Q}) \phi_C - (\varepsilon_0 \dot{H}_i^* N_i + e N_a \dot{R}^+) \phi_F \, dA \quad (103)$$

$$- \int_{A_{Ae}} M_{A+}(\dot{R}^+ + j^+ + \dot{P}_A) + M_{AR}(\dot{R} - j_A - \dot{P}_A) + M_{AQ}(\dot{Q} - i_{Ae} + \dot{P}_A) dA$$

$$- \int_{A_{0Ce}} M_{C+}(\dot{R}^+ + j^+ + \dot{P}_C) + M_{CR}(\dot{R} - j_C - \dot{P}_C) + M_{CQ}(\dot{Q} - i_{Ce} + \dot{P}_C) dA$$

where $\partial G^A / \partial t$, $\partial G^C / \partial t$ and $\partial G^F / \partial t$ are defined in eqs. (29)-(31). Next, note that the kinematic relations (9) and (11) at the anode/electrolyte interfaces imply that

$$\int_{A_{Ce}} (\tilde{\sigma}_{ij}^C \dot{x}_i^C - \tilde{\sigma}_{ij}^F \dot{x}_i^F) n_j \, dA + \int_{A_{Ae}} (\tilde{\sigma}_{ij}^A \dot{x}_i^A - \tilde{\sigma}_{ij}^F \dot{x}_i^F) n_j \, dA$$

$$= \int_{A_{Ce}} (\tilde{\sigma}_{ij}^C \dot{x}_i^C - \tilde{\sigma}_{ij}^F \dot{x}_i^C - \tilde{\sigma}_{ij}^F \dot{F}_{ik}^* a N_k) n_j \, dA + \int_{A_{Ae}} (\tilde{\sigma}_{ij}^A \dot{x}_i^A - \tilde{\sigma}_{ij}^F \dot{x}_i^A - \tilde{\sigma}_{ij}^F \dot{F}_{ik}^* a N_k) n_j \, dA \quad (104)$$

$$+ \int_{A_{Ce}} \tilde{\sigma}_{ij}^F \frac{\partial x_i^C}{\partial X_k} N_k \Omega N_a j_C - \tilde{\sigma}_{ij}^F \frac{\partial x_i^F}{\partial X_k} N_k \Omega N_a j_C) n_j \, dA + \int_{A_{Ae}} (-\tilde{\sigma}_{ij}^F \frac{\partial x_i^F}{\partial X_k} N_k \Omega N_a j_A) n_j \, dA$$

Using Nansen's formula $dA n_i = J F_{ki}^{-1} N_k dA_0$ the last two lines of this result can be rewritten as

$$\int_{A_{Ce}} \tilde{\sigma}_{ij}^F \frac{\partial x_i^C}{\partial X_k} N_k \Omega_C N_a j_C - \tilde{\sigma}_{ij}^F \frac{\partial x_i^F}{\partial X_k} N_k \Omega N_a j_C^+ )n_j dA + \int_{A_{Ae}} (-\tilde{\sigma}_{ij}^F \frac{\partial x_i^F}{\partial X_k} N_k \Omega N_a j_A^+ )n_j dA$$

$$= \int_{A_{0Ce}} \tilde{\sigma}_{ij}^F \frac{\partial x_i^C}{\partial X_k} N_k \Omega_C N_a j_C - \tilde{\sigma}_{ij}^F \frac{\partial x_i^F}{\partial X_k} N_k \Omega N_a j_C^+ )J F_{lj}^{-1} N_l dA_0 + \int_{A_{0Ae}} (-\tilde{\sigma}_{ij}^F \frac{\partial x_i^F}{\partial X_k} N_k \Omega N_a j_A^+ )J F_{lj}^{-1} N_l dA_0$$

Substituting these results back in (103) and collecting terms yields the required result.

**APPENDIX B: List of symbols**

| | |
|---|---|
| $A$ | Width of Stern layer |
| $A_{0Aext}$ ($A_{Aext}$) | Portion of undeformed (deformed) anode surface subjected to traction |
| $A_{0Ae}$, $A_{0Ce}$ | Surface of undeformed anode (cathode) in contact with electrolyte |
| $C$ | Molar concentration of Li in anode |
| $C^{\pm}$ | Molar concentration of activated complex per unit reference area in reaction layer |
| $C_{ijkl}$ | Tensor of elastic moduli |
| $D_{ij}^p$ | Plastic stretch rate tensor |
| $D_q$, $D$ | Mobility of electrons and Li atoms in anode |
| $D_+$ | Mobility of Li$^+$ in the electrolyte |
| $\dot{d}_0$ | Characteristic plastic strain rate in viscoplastic flow potential |
| $E$ | Magnitude of electron charge |
| $E$ | Young's modulus of Si film |
| $E_{ij}^e$ | Elastic Lagrange strain |
| $F_{ij}$ | Deformation gradient |
| $F_{ij}^e$ | Elastic part of deformation gradient |
| $F_{ij}^p$ | Plastic part of deformation gradient |
| $F_{ij}^*$ | Deformation gradient inside double layer |
| $H_i$ | Nominal electric field vector in undeformed solid |
| $H_i^*$ | Electric field vector inside double layer |
| $H_i^A$, $H_i^F$, $H_i^C$ | Electric fields just outside the Stern layers in the anode, electrolyte and cathode |
| $I_A$ | Current flow from anode into the interface per unit reference area |
| $I_0$ | Exchange current density |
| $I^*$ | Electric current flow per unit area from anode to cathode. |
| $i_i$ | Molar flux of electrons per unit reference area in anode |
| $i_{Ae}$, $i_{Ce}$ | Molar flux of electrons per unit reference area flowing from anode (A) or cathode (C) into reaction layer |
| $j_i$ | Molar flux of Li per unit reference area in anode |
| $j_{Ae}$, $j_{Ce}$ | Molar flux of Li per unit reference area flowing from anode or cathode into reaction layer |
| $j_A^+$, $j_C^+$ | Molar flux of Li ions per unit reference area flowing from reaction layer to electrolyte at anode or cathode |
| $J$ | Jacobian of deformation gradient det($\mathbf{F}$) |
| $K_+$, $K_-$ | Forward and reverse reaction rate constants for reaction in reaction layer |
| $k_+, k_-$ | Reaction rate coefficients expressed in terms of bulk concentrations |
| $L_{ij}$ | Velocity gradient |
| $M$ | Stress exponent in viscoplastic flow potential |
| $M_{AR}, M_{A+}, M_{AQ}$ | Chemical potentials of Li atoms, Li$^+$, and electrons in the Stern layer at anode |
| $M_{CR}, M_{C+}, M_{CQ}$ | Chemical potentials of Li atoms, Li$^+$, and electrons in the Stern layer at cathode |
| $M^\pm$ | Electrochemical potential of activated complex in reaction layer |
| $N_a$ | Avogadro number |
| $n_i$ | Normal to deformed anode or cathode surface |
| $N_i$ | Normal to undeformed anode or cathode surface |
| $\dot{P}_A, \dot{P}_C$ | Rate of reaction, in mols of Li per second per reference area, at anode or cathode. |
| $q$ | Molar density of mobile electrons per unit reference volume of anode |
| $q_0$ | Concentration of free electrons in electrically neutral solid |
| $Q$ | Molar concentration of free electrons per unit reference area in reaction layer |
| $R$ | Molar concentration of Li per unit reference area in reaction layer |
| $R^+$ | Molar concentration of Li ions per unit reference area in reaction layer |
| $\bar{R}$ | Gas constant |
| $S_{ijkl}$ | Elastic compliance tensor of the anode |
| $T$ | Temperature |
| $t_i$ | Traction acting on external solid surface |
| $W_{ij}^p$ | Plastic spin tensor |
| $V_{0A}, V_{0C}$ | Volume occupied by undeformed anode |
| $x_i$ | Position vector of material point in deformed system |
| $X_i$ | Position vector of material point in undeformed system |
| $\alpha$ | Symmetry factor in Butler-Volmer equation |
| $\beta$ | Volume expansion ratio of anode when lithiated |
| $\Gamma, \Gamma^+, \Gamma_Q$ | Activity coefficients of Li atoms, Li$^+$ and electrons in reaction layer |
| $\gamma, \gamma_q, \gamma^+$ | Activity coefficients of Li, electrons and Li$^+$ just outside Stern layer |
| $\varepsilon_0$ | Permittivity of free space |
| $\varepsilon$ | In-plane elastic strain in Si film |
| $\phi$ | Electric potential |
| $\Delta\phi_0$ | Rest potential for reaction layer |
| $\Delta\phi_{AF}$ | Overpotential for reaction layer at anode |
| $\psi^A, \psi^F, \psi^C$ | Free energy per unit undeformed volume in the anode, electrolyte and cathode |
| $\Psi^A, \Psi^C$ | Free energies per unit reference area of the Stern layer at the anode and cathode |
| $\psi_0^A(c,q)$ | Free energy per mol of reference lattice for a stress and charge free anode |
| $\psi_0^F(\rho^+)$ | Free energy per mol of stress free and electrically neutral electrolyte |
| $\eta_i$ | True electric field vector in deformed solid or electrolyte |
| $\lambda_p$ | Out-of-plane plastic stretch in Si film |
| $\mu_A$ | Chemical potential of interstitial Li |
| $\mu_{Aq}, \mu_{Cq}$ | Electrochemical potential of free electrons in the anode and cathode |
| $\mu^+$ | Electrochemical potential of Li$^+$ in the electrolyte |
| $\pi$ | Pressure in electrolyte |
| $\pi_{ij}^*$ | Maxwell pressure inside Stern layer |
| $\rho$ | Molar density of Li per unit reference volume of anode |
| $\rho_{0A}$ | Molar density of Si per unit reference volume in anode |
| $\rho^+$ | Molar density of Li$^+$ per unit reference volume in electrolyte |
| $\rho_0^+$ | Concentration of Li$^+$ in an electrically neutral electrolyte |
| $\sigma_0$ | Yield stress |
| $\sigma_{ij}$ | True (Cauchy) stress |
| $\sigma_{ij}^M$ | Maxwell stress |
| $\tilde{\sigma}_{ij}^A, \tilde{\sigma}_{ij}^F, \tilde{\sigma}_{ij}^C$ | Combined mechanical and Maxwell stresses in the solid adjacent to the electrode/electrolyte interfaces in the anode, electrolyte, and cathode |
| $\tau_{ij}^D$ | Deviatoric Kirchhoff stress |
| $\tau_e$ | Von-Mises effective Kirchhoff stress |
| $\Omega$ | Volume of a single Li atom in the un-deformed cathode |

**TABLE 1: Parameters used in modeling thin film electrode. Parameters without cited references were fit to results shown in Fig 3.**

| Thickness of unlithiated Si electrode $h_{A0}$ | 250 nm (Sethuraman *et al.* 2010c) |
|---|---|
| Molar density of Si $\rho_{0A}$ | 7.874 $10^4$ mol/m$^3$ (Mohr *et al.* 2008) |
| Mass density of Si | 2.2g cm$^{-3}$ |
| Modulus of unlithiated Si electrode $E_0$ | 100 GPa (Sethuraman *et al.* 2010d) |
| Rate of change of Si electrode modulus with concentration $\partial E / \partial c$ | 20 GPa |
| Poisson's ratio of Si electrode $\nu$ | 0.26 |
| Volumetric expansion of Si with Li concentration ($\partial \beta / \partial c$) | 0.7 [Obrovac and Krause, 2007] |
| Characteristic strain rate for plastic flow in Si ($\dot{d}_0$) | 0.8 $10^{-9}$ s$^{-1}$ |
| Initial yield stress of Si ($\sigma_0$) | 0.12 GPa |
| Rate of change of flow stress in Si with concentration ($\partial \sigma_0 / \partial c$) | 0.03 GPa |
| Stress exponent for plastic flow in Si $m$ | 4 |
| Initial residual stress in Si film | 0.25 GPa |
| Initial concentration of Li in Si | 0.0078 |
| Initial concentration of Li$^+$ in Stern layer $R_0^+$ | 0.01 mol/m$^2$ |
| Stress free rest potential at reference concentration ($\bar{R}T / eN_a)\Theta(c_0)$ | 0.78V |
| Rate of change of rest potential with concentration ($\bar{R}T / eN_a)\partial \Theta / \partial c$ | -0.16V |
| Reaction rate constant at anode $k_A$ | 2 A m$^{-2}$ |
| Symmetry factor $\alpha$ | 0.5 |
| Thickness of Stern layer ($a$) | 1 nm |
| Permittivity of free space ($\varepsilon_0$) | 8.85 x $10^{-12}$ (Mohr *et al.* 2008) |
| Faraday constant ($eN_a$) | 96485 C/mol (Mohr *et al.* 2008) |
| Gas constant ($\bar{R}$) | 8.314 J K$^{-1}$ mol$^{-1}$ (Mohr *et al.* 2008) |
| Temperature ($T$) | 298K |